\documentclass{nature}
\usepackage{graphicx}
\usepackage{amsmath}
\usepackage{booktabs}
\usepackage{placeins}
\usepackage{amssymb}
\usepackage{textcomp}
\usepackage{gensymb}
\usepackage{xcolor}
\usepackage{lineno}
\usepackage{soul}
\usepackage{colortbl}
\usepackage{ulem}
\usepackage[labelfont=bf]{caption}
\usepackage{multibib}
\makeatletter
\let\saved@includegraphics\includegraphics
\usepackage{float}
\AtBeginDocument{\let\includegraphics\saved@includegraphics}
\renewenvironment*{figure}{\@float{figure}}{\end@float}
\makeatother
\newcommand{\reftextit}[1]{}
\usepackage{appendix}
\usepackage{bibunits}

\usepackage{graphicx}

\newcommand{\ourtitle}{
Quasi-one-dimensional taco-shaped bands in large-angle twisted bilayer transition metal dichalcogenides
}

\title{\ourtitle}

\author{
 Giovanny Espitia$^{1*}$,
 Seung Hun Lee$^{1*}$, Calvin Kaiyu Chiu$^{1}$, Junyeong Ahn$^{1 \dagger}$ and Mit H. Naik$^{1 \dagger}$
}

\date{\today} 

\begin{document}

\maketitle 
\begin{affiliations}
 \item Department of Physics and Center for Complex Quantum Systems, The University of Texas at Austin, Austin, Texas, 78712, USA.
\end{affiliations}
\noindent 
*these authors contributed equally\\
$\dagger$ Corresponding authors:~\textcolor{blue}{{junyeong.ahn@austin.utexas.edu,~mit.naik@austin.utexas.edu}}

\newcommand{\MHN}[1]{{\color{orange}(MHN)#1}}

\begin{abstract}
Two-dimensional moiré materials offer a powerful, twist-tunable platform for engineering electronic bands and correlations, though most studies to date have focused on small twist angles where flat bands arise from symmetry-pinned monolayer momenta.
Here, we observe the surprising emergence of flat electronic bands with a distinctive quasi-one-dimensional dispersion at large twist angles in bilayer transition metal dichalcogenides that originate from the $\Lambda$ valley states at generic momenta between $\Gamma$ and $K$ points.
These {\it taco-shaped} anisotropic bands result from optimal interlayer hybridization between like-spin $\Lambda$ valleys at the conduction band minimum in the Brillouin zone, resulting in directional band flattening at a magic twist-angle of 21.8\(^{\circ}\). The bands form six anisotropic channels with a sixfold alternating spin texture reminiscent of altermagnetic textures.
At low energies, the density of states shows a power-law dependence due to
the quasi-one-dimensional character, enhancing the potential for correlated phases. Our results provide a new platform for correlated phenomena and broaden the scope of moiré engineering to large twist angles in 2D materials.

\end{abstract}

\renewcommand{\baselinestretch}{1.7}
\setlength{\parskip}{7pt}

\begin{bibunit}
Materials whose electronic dispersion contains flat bands are interesting due to their ability to host correlated phases of matter that lead to phenomena such as unconventional superconductivity \cite{cao2018unconventional, cao2018correlated, li2024imaging, xia2024unconventional, zhou2021superconductivity}, quantum anomalous Hall effect \cite{cai2023signatures, park2023observation, zeng2023thermodynamics, lu2024fractional, li2021quantum}, and generalized Wigner crystalizations \cite{li2021imaging, regan2020mott, li2024wigner}. A versatile way to engineer such bands is to form a moiré superlattice by rotating one two-dimensional layer relative to another \cite{carr2017twistronics}.  The prototypical case is twisted bilayer graphene, where a series of small “magic” angles generate flat bands near the Fermi level \cite{bistritzer2011moire, balents2020superconductivity, carr2019exact}. More recently, twisted bilayers of transition-metal dichalcogenides (TMDs) have exhibited analogous correlated behavior at small twist angles  \cite{devakul2021magic, naik2018ultraflatbands, liu2021excitonic}. So far, most studies have focused on flat bands \cite{li2021imagingmoire}, correlated states \cite{li2021imaging,regan2020mott,li2024imagingtunable,cai2023signatures, xia2024unconventional, xu2020correlated, wu2018hubbard}, and novel moiré excitons \cite{tran2019evidence, naik2022intralayer, alexeev2019resonantly, susarla2022hyperspectral,li2024imaging,hu2023light} at small twist angles in transition metal dichalcogenides driven by structural reconstructions and inhomogeneous interlayer hybridization in large-area moiré superlattices \cite{naik2018ultraflatbands, naik2020origin}.
Despite these rich phenomena, a common limitation of small-angle moiré superlattices is their strong sample heterogeneity \cite{turkel2022orderly,li2024imagingtunable} and sensitivity to sample preparation. This arises from their tendency toward domain-wall formation and disorder \cite{turkel2022orderly, tillotson2024scanning, weston2020atomic,li2024imagingtunable}, induced by substantial structural reconstructions driven by the strongly varying generalized stacking-fault energy landscape~\cite{carr2018relaxation, enaldiev2020stacking}.
In contrast, large-angle moiré superlattices are more robust against layer misalignment and manifest less disorder due to reduced structural relaxation \cite{maity2020phonons, li2025robust, li2024tuning}, offering a promising route to combine enhanced stability with the tunability of moiré engineering.

Large twist angles in TMDs, however, have received little attention, since flat bands do not arise from simple moiré hybridization of high-symmetry-point valleys in this regime. The most studied valley in transition metal dichalcogenides is at the $K$ point in the Brillouin zone. The valence and conduction band extrema at the $K$ valleys originate primarily from transition-metal $d$ orbitals, resulting in wavefunctions that are highly confined within individual layers and exhibit weak interlayer hybridization \cite{li2021imaging, naik2017origin, bradley2015probing}. At large twist angles, the substantial momentum mismatch between the $K$ valleys of the two layers leads to nearly layer-degenerate states at the band edges and small gap openings appear at valley crossing points at energies farther from the valence and conduction band edges.  At small twist angles, the $\Gamma$ valleys in the valence band exhibit significant interlayer hybridization and are strongly affected by the varying interlayer spacing in the superlattice \cite{naik2020origin, angeli2021gamma}. On the other hand, at large twist angles, since the interlayer spacing shows no variation, the $\Gamma$ valleys do not host any flat bands. The emergence of flat bands at large twist angles in bilayers has only recently been realized using approaches that go beyond conventional moiré superlattice formation, employing super-moiré structures \cite{li2025robust}, namely, large-scale moiré-of-moiré structures formed at small deviations from commensurate angles.


In this work, we demonstrate a new mechanism for the emergence of a flat-band dispersion in a class of large-angle TMDs, realized without relying on lattice reconstructions.
In most natural bilayer TMDs (MoS$_2$, MoSe$_2$, WS$_2$, WSe$_2$), the conduction band minimum (CBM) lies at the $\Lambda$ valley of the Brillouin zone, located approximately midway between the high-symmetry $\Gamma$ and $K$ points \cite{bradley2015probing,splendiani2010emerging}. The conduction-band wavefunctions at the $\Lambda$ valley exhibit stronger interlayer hybridization compared to the valence and conduction band $K$ valleys owing to their substantial out-of-plane extent, arising from dominant Se-$p_z$ orbital character \cite{naik2017origin,splendiani2010emerging}. In this work, we investigate the emergent electronic phenomena that arise when the relative separation between these valleys is tuned by twisting the layers. Our first-principles calculations and theoretical model reveal the emergence of a unique flat band with a taco-shaped quasi-one-dimensional dispersion in large-angle twisted bilayer TMDs. The flat band originates from optimal interlayer hybridization between like-spin $\Lambda$ valleys at the CBM of the Brillouin zone. The emergent flat band consists of six elongated valleys connected across the moiré Brillouin zone, forming quasi-one-dimensional electronic pathways with alternating spin polarization under sixfold rotation, reminiscent of altermagnetic spin textures. The reduced dimensionality of these bands produces a power-law divergent density of states, potentially promoting strong electronic interactions and correlated behavior.

This band flattening is driven by a delicate interplay between the interlayer coupling strength and the spin texture of adjacent $\Lambda$ valleys of opposite layers, which align optimally only at the specific twist angle of 21.8$^{\circ}$. At this "magic" angle, the interlayer hybridization is sufficiently strong as well as spin-compatible to produce a remarkably flat band connecting the adjacent $\Lambda$ valleys in the Brillouin zone (BZ). Even small departure from this angle rapidly disrupts the alignment of the $\Lambda$ valleys, removing the conditions necessary for flat-band formation.
While strong interlayer coupling through Umklapp scattering at commensurate twist angles, such as 21.8$^{\circ}$, has been reported in graphene bilayers \cite{bistritzer2010transport}, we emphasize that the magic angle here does not arise solely from enhanced hybridization.

\section*{Directional band flattening at 21.8$^{\circ}$ twist}
\begin{figure}[ht]
 \includegraphics[width=1.\linewidth
 ]{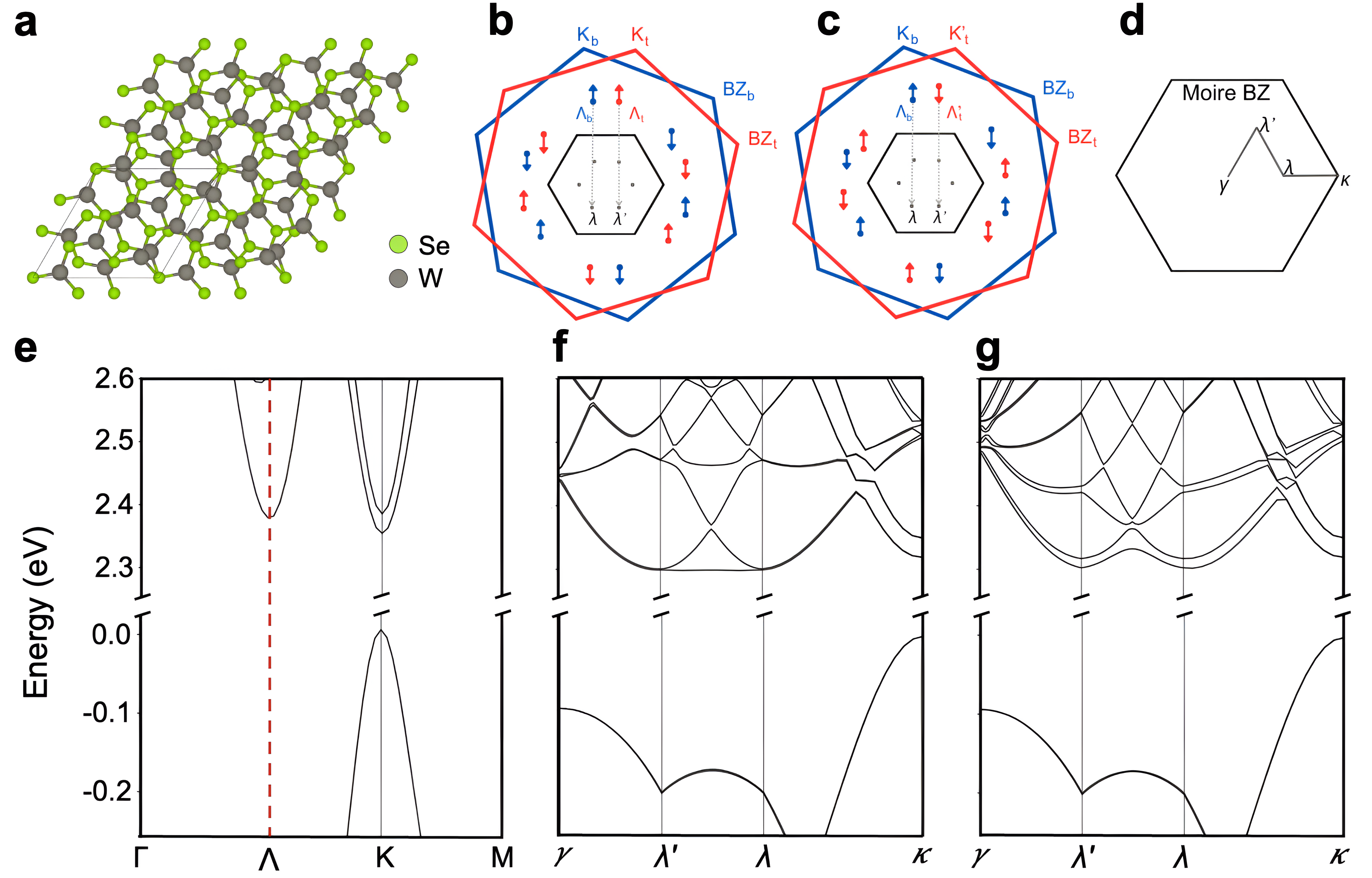}
    \caption{
    \textbf{Electronic structure of twisted WSe$_2$ at 21.8$^{\circ}$ versus 38.2$^{\circ}$}. \textbf{(a)} 21.8$^\circ$ twisted WSe$_2$ structure, \textbf{(b), (c)} The red (blue) hexagon is the unit-cell BZ for the top (bottom) layer twisted at 21.8$^{\circ}$ and 38.2$^{\circ}$, respectively. The red (blue) arrows represent spins of the conduction band minimum states at the $\Lambda$ valleys for the top and bottom layers, respectively.  \textbf{(d)} The moiré BZ with the path used to plot the band structures in \textbf{(f)} and \textbf{(g)} for 21.8$^{\circ}$ and 38.2$^{\circ}$ t-WSe$_2$, respectively. \textbf{(e)} shows the band structure of monolayer WSe$_2$.
    }
    \label{fig:fig1}
\end{figure}

First, we describe the origin of the flat bands that emerge from the $\Lambda$ valley. Fig.~\ref{fig:fig1}a shows the simulated 21.8$^{\circ}$ moiré unit cell in real space. 
At this commensurate twist angle, the resulting moiré superlattice is a $\sqrt{7}\times\sqrt{7}$ supercell relative to the primitive unit cell.
The individual BZs of the top and bottom WSe$_2$ layers are shown in Fig.~\ref{fig:fig1}b and Fig.~\ref{fig:fig1}c; where the central smaller hexagon is the common moiré BZ onto which the individual layer k-points fold into at a twist angle of $21.8^{\circ}$ and $38.2^{\circ}$, respectively. In the monolayer, the $\Lambda$ point in the BZ is defined as the valley occurring roughly half-way along the $\Gamma-K$ line (Fig.~\ref{fig:fig1}e). The $21.8^{\circ}$ and $38.2^{\circ}$ are mirror-symmetric and have the same size superlattice, but different proximity of the top and bottom layer valleys. In $21.8^{\circ}$, the bottom layer's $\Lambda_b$ valley is closer to the top layer's $\Lambda_t$ valley with the same spin, while in $38.2^{\circ}$, the $\Lambda_b$ valley is closer to the $\Lambda'_t$ valley with the opposite spin (Fig.~\ref{fig:fig1}b and c). We expect a large gap opening at the k-point where the $\Lambda$ valleys of the two layers intersect with the same spin.

Fig.~\ref{fig:fig1}d shows the path in the BZ used to compute the band structures.Fig.~\ref{fig:fig1}f displays the band structure computed using density functional theory (DFT) calculations with $GW$ (where G stands for the single-particle Green's function and W is the screened Coulomb interaction) self-energy corrections \cite{hybertsen1986electron} for both the 21.8$^{\circ}$ twisted WSe$_2$ and the mirror-symmetric 38.2$^{\circ}$ structure. Along the path between the $\lambda$ and $\lambda^{\prime}$ points, a flat band emerges only in the 21.8$^{\circ}$ superlattice driven by strong hybridization between the like-spin adjacent valleys of the two layers (Fig.~\ref{fig:fig1}b). On the other hand, in the 38.2$^{\circ}$ twisted bilayer, opposite spins in the adjacent valleys of the two layers (Fig.~\ref{fig:fig1}b) couple weakly, restoring a dispersive conduction band. In the 21.8$^{\circ}$ system, we also observe a second flat band emerge along the same path in the BZ at an energy separation of  $\sim$150 meV from the CBM. This is related to a similar coupling between the higher energy valleys at the $\Lambda$ point separated through spin-orbit coupling (see comparison to individual monolayers in SI Fig.~S1). 
As expected, the valence and conduction band valleys at the $\kappa$ point in the moiré BZ do not feature any renormalization due to the large twist angle. In 21.8$^{\circ}$, the VBM is two-fold degenerate and in 38.2$^{\circ}$, Umklapp scattering induces a small gap of 2.7 meV. 

\begin{figure}[ht]
 \includegraphics[width=1.\linewidth]{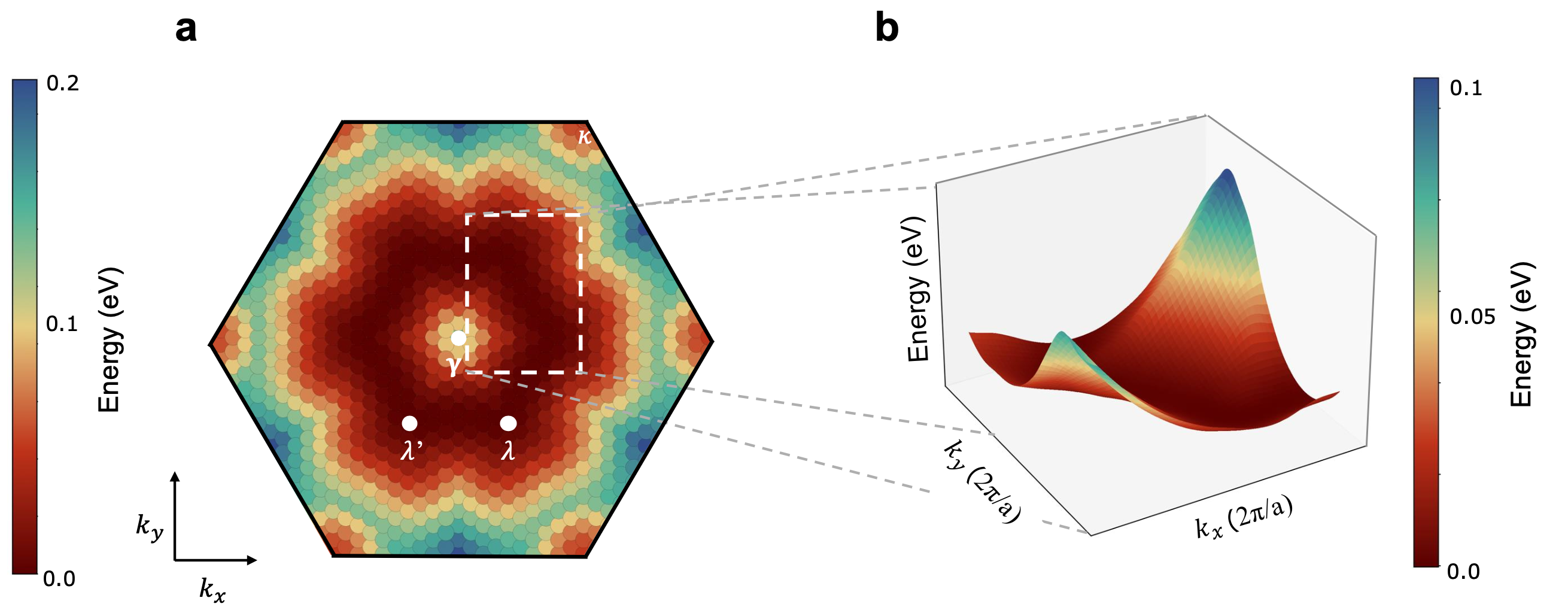}
    \caption{\textbf{Energy distribution of the conduction band edge in twisted WSe$_2$ at 21.8$^{\circ}$}. The energy is relative to the conduction band minimum (fixed at 0 eV). On the right, we show a section of the energy distribution in 3D which connects two $\lambda$ points in the moir\'e BZ.}
    \label{fig:fig2}
\end{figure}

Fig.~\ref{fig:fig2}a visualizes the energy of the first unoccupied band in the moiré BZ for the 21.8° structure. The band has a strong anisotropy at the conduction band minimum, forming a flat band connecting the six $\lambda$ points in the BZ. Moving away from this flat valley in a perpendicular direction causes the energy to rise sharply as displayed in Fig.~\ref{fig:fig2}b. We refer to this unique quasi-one-dimensional valley as a taco flat band (Fig.~\ref{fig:fig2}b).  We note that this dispersion is fundamentally different from other well-known band structures such as the Mexican hat dispersion which features a central energy peak surrounded by an isotropic ring of low-energy states. Additionally, this dispersion is distinct because each segment connecting the adjacent $\lambda$ valleys has a unique spin. 


\section*{Alternating spin structure}
\begin{figure}[ht]
 \includegraphics[width=1.\linewidth]{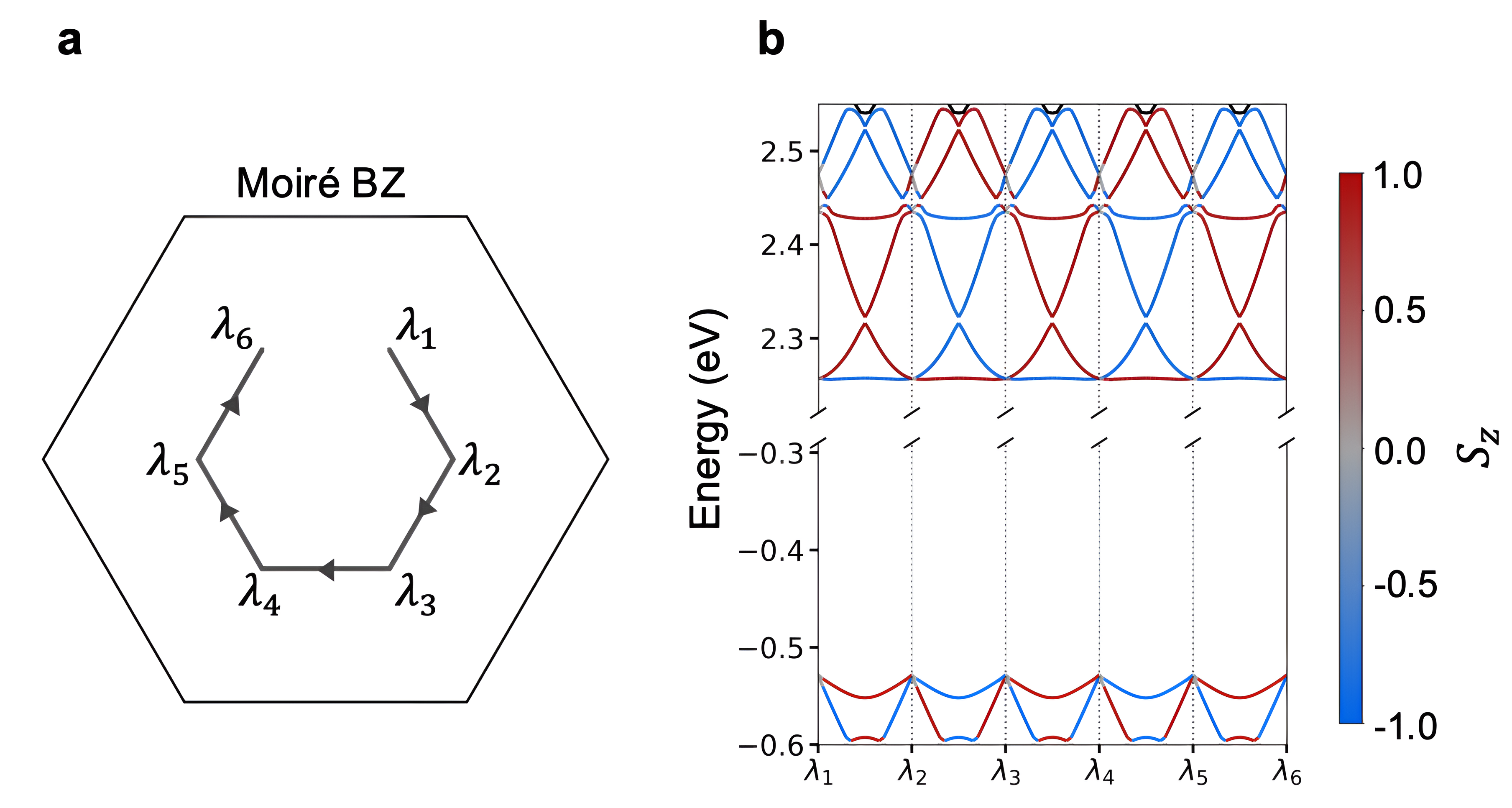}
    \caption{\textbf{Alternating spin texture of the flat bands}. \textbf{(a)} High-symmetry path  within the moir\'e BZ used for the band structure calculation. \textbf{(b)} Spin-resolved band structure along the specified path. The color bar indicates the expectation value of the out-of-plane spin component, $\langle S_z \rangle$, with red corresponding to spin-up and blue to spin-down. Grey regions denote a vanishing net spin polarization where the bands are degenerate. The path highlights the alternating spin polarization of the flat bands in adjacent $\lambda$ valleys.}
    \label{fig:fig3}
\end{figure}

To examine the spin texture of the 21.8$^{\circ}$ twisted bilayer WSe$_2$, we computed the expectation value of the out-of-plane spin operator, $\langle S_z \rangle$, for each electronic state (more details in the supplementary information). Traversing the path connecting $\lambda$ valleys in the moiré BZ (Fig.~\ref{fig:fig3}a), the spin character of the flat bands alternates (Fig.~\ref{fig:fig3}b).  
This unique spin texture is a direct signature of the underlying spin-selective interlayer hybridization and it represents a key feature of the taco flat band. The spin alternates for every 60° rotation, which is reminiscent of $f$-wave altermagnetism \cite{vsmejkal2022emerging}. The higher energy flat band, which is a result of spin-orbit coupling, has the opposite spin.


\section*{Evidence of magic angle at 21.8$^{\circ}$}
\begin{figure}[ht]
 \includegraphics[width=1.\linewidth]{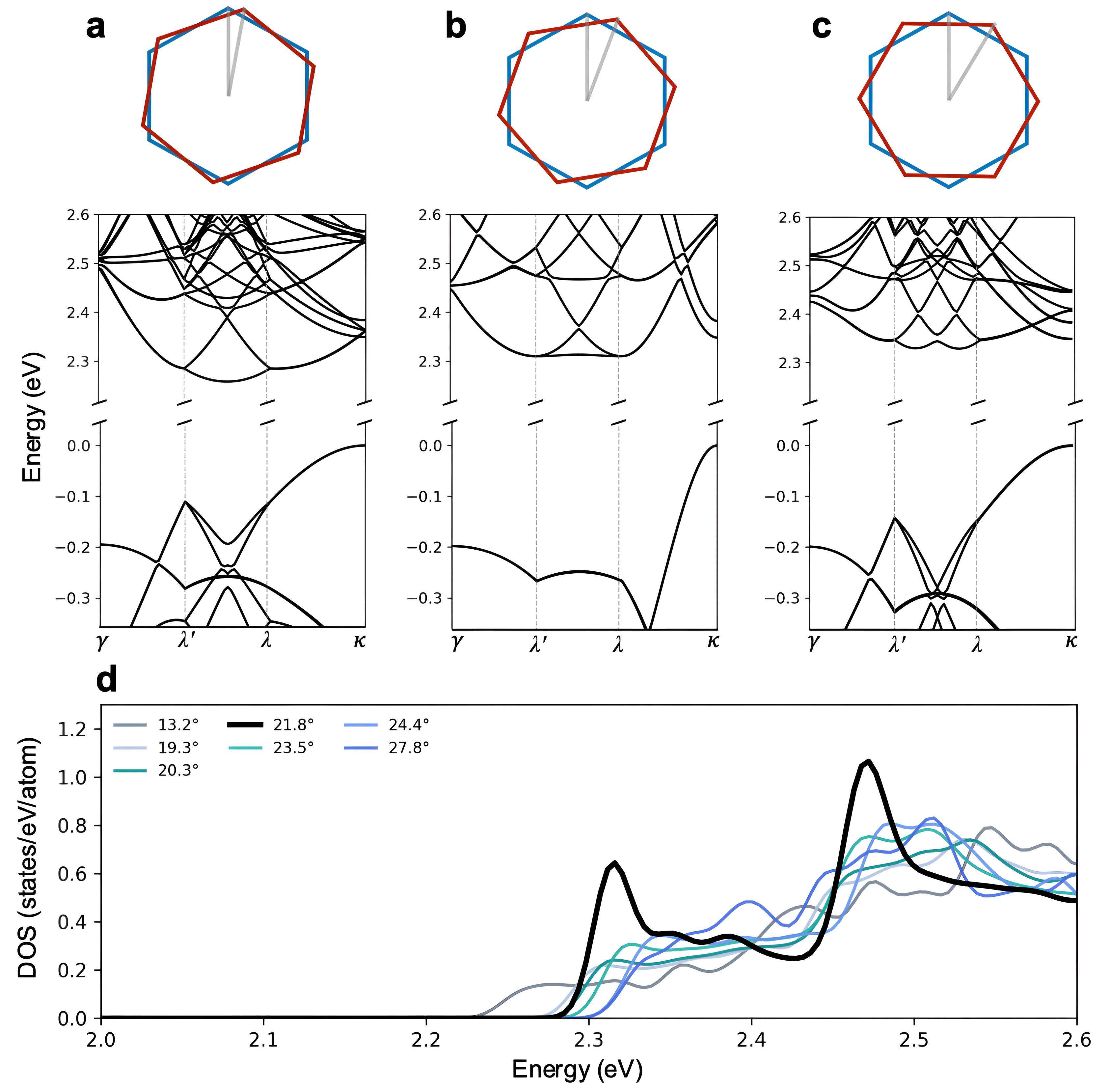}
    \caption{\textbf{Evolution of the electronic band structure and density of states (DOS) with twist angle}. The band structure of twisted bilayer WSe$_2$ is shown for three commensurate twist angles, calculated along the high-symmetry path $\gamma$-$\lambda'$-$\lambda$-$\kappa$. \textbf{(a)} At $13.17^{\circ}$, the conduction band minimum (CBM) shows a pronounced downward curvature due to strong interlayer hybridization. \textbf{(b)} At the magic-like angle of $21.8^{\circ}$, the CBM becomes notably flat between the $\lambda$ and $\lambda'$ valleys. \textbf{(c)} At $27.8^{\circ}$, weaker hybridization restores a more conventional, upward-curving dispersive band. \textbf{d)} The plot shows calculated DOS for twisted bilayer WSe$_2$ at various commensurate angles. The VBM for each structure is subtracted in its corresponding curve. A sharp peak emerges near the conduction band minimum only at the magic-like angle of $21.8^{\circ}$. For other angles, this feature broadens and its amplitude is reduced. The second, smaller peak for the $21.8^{\circ}$ structure arises from the second flat band, which has the opposite spin polarization.}
    \label{fig:fig4}
\end{figure}

To understand the electronic structure as one deviates from the 21.8$^{\circ}$ angle, we performed DFT calculations on six additional twisted WSe$_2$ structures. The range of angles considered was between 13.17$^{\circ}$ to 27.8$^{\circ}$. We corrected the DFT band gap by a rigid shift to the GW band gap of natural bilayer WSe$_2$. In Fig.~\ref{fig:fig4}b, we show the electronic band structure at 21.8$^{\circ}$. At this angle, the distance in momentum space between adjacent $\lambda$ valleys is optimal such that the interlayer hybridization induces a flat band along the $\lambda - \lambda^{\prime}$ path.  Figures~\ref{fig:fig4}a and \ref{fig:fig4}c illustrate the limiting cases. At $13.17^{\circ}$, which is a commensurate $\sqrt{19} \times \sqrt{19}$ size superlattice, the smaller momentum difference between $\lambda$ valleys reduces the energy at which the valleys cross and the hybridization leads to a dispersive band with a pronounced negative curvature. In contrast, at $27.8^{\circ}$, which is a commensurate $\sqrt{13} \times \sqrt{13}$ size supercell, the larger valley separation leads to a dispersive band connecting the $\lambda$ valleys with positive curvature. In Fig.~\ref{fig:fig4}d, we show the DOS for all of the structures considered. The DOS further highlights the unique behavior at $21.8^{\circ}$. Specifically, only at this angle does a vHS appear at the CBM. This vHS is a clear signature of the flat band formed between the $\lambda$ valleys. Moving away from $21.8^{\circ}$ broadens the peak and reduces its amplitude. An apparent ordering of the peaks is present according to the twist angle. This successive ordering is a direct result of how the twist angle tunes the band structure. The proximity of adjancent $\lambda$ valleys at the 13.17$^{\circ}$ twist angle pushes the CBM to a lower energy, causing the DOS peak to appear first. At the 27.8$^{\circ}$ twist angle, the interlayer hybridization is weaker and thus the CBM relaxes to a higher energy leading to a DOS peak that appears last. The "magic" angle, 21.8$^{\circ}$, represents an intermediate case. A second large DOS peak at higher energy in $21.8^{\circ}$ arises from the second flat band, described above. Twist angles within $\pm0.5^{\circ}$ of $21.8^{\circ}$ require superlattice calculations with more than 55600 atoms and therefore prohibitively large for DFT calculations; instead, we construct a continuum model fitted to the DFT data to gain a richer understanding of the twist-angle dependence and quasi-1D physics of the taco flat band.

\section*{Magic angle near 21.8$^{\circ}$ in the low-energy model}

    We analyze the twist-angle dependence near 21.8$^{\circ}$ using a simplified low-energy model for the conduction bands (see Supplementary Information for details). By folding the $\Lambda$ valleys of the lowest conduction bands of the top and bottom layers into the moiré Brillouin zone (mBZ), we obtain a $12\times12$ Hamiltonian that captures the relevant low-energy physics.

    To describe the band flattening along the $\lambda$–$\lambda'$ line, we focus on a $2\times2$ subblock of the full Hamiltonian describing a pair of adjacent $\lambda$ and $\lambda'$ valleys. Taking the midpoint of the $\lambda$–$\lambda'$ line as the origin, the effective Hamiltonian reads
    \begin{equation}\label{maineffH}
        H(k)=
        \begin{pmatrix}
            \frac{(k-k_{0})^2}{2m} & t(k) \\
            t^*(k) & \frac{(k+k_{0})^2}{2m}
        \end{pmatrix},
    \end{equation}
    where $k_{0}=r_{\textrm{shift}}\times(2\pi/3)\sin(\theta/2)$ $a_0^{-1}$, with $r_{\textrm{shift}}\approx0.7$ denoting the relative shift of the valleys in twisted bilayer WSe$_2$ compared to the monolayer, $a_0$ the monolayer lattice constant, $m\approx0.80177~a_0^{-2}$ eV$^{-1}$ the effective mass, and $k$ the one-dimensional momentum.
    Because of the phenomenological moiré potential and the intrinsic anisotropy of the $\Lambda$ valleys, both $r_{\textrm{shift}}$ and $m$ depend on $\theta$. However, the $\theta$-dependence of $k_0$ alone already captures the essential features of the magic angle. We therefore first consider a simplified model where only $k_0$ varies with $\theta$, keeping $r_{\textrm{shift}}$ and $m$ fixed, and later compare it with full numerical results that include all $\theta$-dependent terms.
    
    Owing to twofold rotational symmetry about the $\gamma$–$\kappa$ ($\gamma$–$\kappa'$) axis and time-reversal symmetry, the tunneling term takes the form $t(k)\approx e^{i\phi_o}(w_0+e^{i\phi_r}w_2 k^2)$, where $\phi_o$ and $\phi_r$ represent the overall and relative phases, respectively. The overall phase does not affect the energy spectrum $E_\pm(k)=(k_0^2 m+k^2 m\pm2\sqrt{k_0^2 k^2 m^2+m^4|t(k)|^2})/2m^2$, where $E_-(k)$ corresponds to the lowest conduction band dispersion. Fitting to first-principles results yields $w_0\approx0.14175$ eV, $w_2\approx0.49357~a_0^2$ eV, and $\phi_r\approx1.105$.
    
    The bandwidth along $\lambda$–$\lambda'$ is given by the largest of three characteristic energy differences:
    \begin{align}
        \Delta E_1&=|E_-(0)-E_-(k_0)|,\\
        \Delta E_2&=|E_-(0)-E_{\textrm{min}}|,\\
        \Delta E_3&=|E_-(k_0)-E_{\textrm{min}}|,
    \end{align}
    where $E_{\textrm{min}}$ denotes the minimum---if present---of $E_-(k)$ between $k=0$ and $k=k_0$ (see Fig.~\ref{Figmain}b).
    For small twist angles, the $\lambda$ and $\lambda'$ valleys strongly overlap and form a single minimum at $k = 0$, giving $\Delta E_1$ as the bandwidth. As $\theta$ increases, $E_-(0)$ rises relative to $E_-(k_0)$, reducing $\Delta E_1$. Beyond a critical angle, the curvature at $k=0$ changes sign, producing two symmetric minima between $k=0$ and $\pm k_0$. If $E_-(0)<E_-(k_0)$ ($E_-(0)>E_-(k_0)$), the bandwidth is determined by $\Delta E_3$ ($\Delta E_2$). Since $\Delta E_2$ increases with $\theta$ while $\Delta E_3$ decreases, the total bandwidth $\max(\Delta E_1,\Delta E_2,\Delta E_3)$ reaches its minimum when $\Delta E_2=\Delta E_3$, defining the \textit{magic angle}. Analytically, this condition yields
    \begin{equation}\label{mangle}
        \theta_c=2\arcsin{\Bigg[\frac{3}{\pi r_{\textrm{shift}}}\sqrt{\frac{|m a_0 w_0|-2 m^2 a_0 w_0 w_2\cos\phi_r}{3+ 4m^2 w_2^2}}\Bigg]},
    \end{equation}
    which evaluates to $21.59^\circ$ for the parameters above. Eq.~\eqref{mangle} implies that a magic angle can generically arise in other twisted bilayer TMD systems with different parameters, as long as $\cos\phi_r<\textrm{sgn}(m a_0 w_0)/(2m w_2)$, ensuring that the expression under the square root remains positive.
    Including the $\theta$ dependence of $r_{\textrm{shift}}$ and $m$ refines the numerical value to $\theta_{c,\text{num}}=21.7^\circ$, in close agreement with the analytic result (Fig.~\ref{Figmain}c). Although discrepancies emerge away from the critical angle, both approaches capture the characteristic flattening near 21.8$^\circ$.
    

    \section*{Power-law dependence of the density of states}
\begin{figure}[t!]
        \centering
  	\includegraphics[width=1.\linewidth]{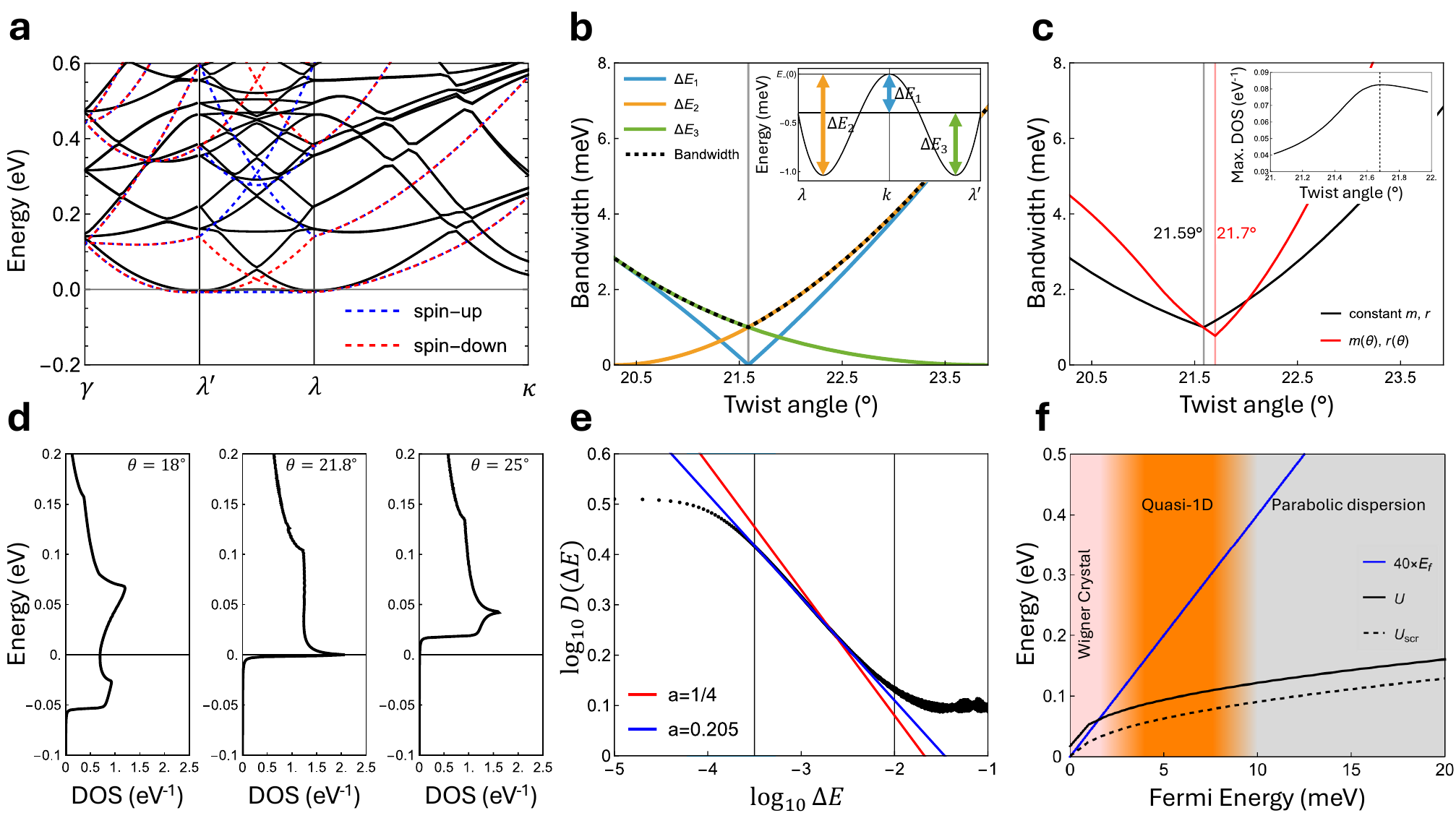}
  	\caption{
        {\bf {Magic twist angle in the continuum model.}}
  		\textbf{(a)} Band structures of spin-up (red dashed) and spin-down (blue dashed) states obtained from the continuum model Hamiltonian. Black solid lines show first-principles results.
        \textbf{(b)} Bandwidth of $E_-(k_y)$, defined as $\max(\Delta E_1, \Delta E_2, \Delta E_3)$ (black dashed line). Inset: $E_-(k)$ at $\theta=21.8^\circ$ with $\Delta E_1$, $\Delta E_2$, and $\Delta E_3$ indicated by blue, orange, and green arrows.
        \textbf{(c)} Comparison of bandwidth from Eq.~\eqref{mangle} (black) and numerical calculations (red). Inset: DOS vs. $\theta$, showing a maximum near $\theta\sim21.7^\circ$.
        \textbf{(d)} DOS $D(E)$ from the continuum model for different twist angles with broadening $\eta=0.5$ meV.
        \textbf{(e)} Log–log plot of $\log_{10} D(\Delta E)$ vs. $\log_{10} \Delta E$ with $\eta=0.1$ meV. Red and blue lines indicate reference power laws with exponents $-1/4$ and $-0.205$. Vertical lines mark the width of the peak ($\sim 10$ meV) and the crossover from the power-law regime to the logarithmic divergence ($\sim 0.3$ meV).
        \textbf{(f)} Coulomb interaction energy (black solid line), gate-screened interaction energy for a sample–gate distance of 5 nm (black dashed line), and 40$\times$ the electronic kinetic energy (blue solid line), all plotted as functions of Fermi energy.
  	}\label{Figmain}
    \end{figure}

    To confirm that the minimized bandwidth at the magic angle leads to the DOS peak observed in DFT calculations, we compute the DOS using the model Hamiltonian.
    Fig.~\ref{Figmain}d shows the calculated DOS for $\theta=18^\circ$, $21.8^\circ$, and $25^\circ$. A pronounced peak appears at the lower edge of the conduction band only at $\theta=21.8^\circ$, originating from the flattening of the taco band dispersion.
    Fig.~\ref{Figmain}e shows the $\log_{10} D(\Delta E)$ vs. $\log_{10} \Delta E$ plot for $\theta=21.8^\circ$, where $\Delta E\equiv E-E_-(0)$ and $E_-(0)$ denotes the divergence energy. The log–log curve is linear in the range $-3.5\lesssim\log_{10} \Delta E\lesssim-2$ with slope $\approx-0.205$, indicating that the DOS follows a power-law relation, $D(\Delta E)\propto (\Delta E)^{-0.205}$. The upper bound of this power-law regime corresponds to the width of the peak structure ($\sim 10$ meV), whereas at higher energies the spectrum crosses over to approximately quadratic dispersion, yielding a constant DOS.
    
    The power-law DOS originates from the effectively quasi-one-dimensional (1D) dispersion. To illustrate this, we consider the model dispersion
    \begin{equation}\label{modiv}
      \varepsilon(\textbf k)=ak_x^2+bk_y^2+ck_y^4.
    \end{equation}
    Several limiting cases are instructive: (i) For $a=b>0$, the dispersion describes a rotationally symmetric band minimum at $k_x=k_y=0$, yielding a constant DOS proportional to the effective mass $1/(2a)$.
    (ii) For $a$ and $b$ of opposite sign, the dispersion forms a saddle point, leading to the well-known vHS with logarithmic divergence at $\varepsilon=0$.
    (iii) For $b=c=0$, the dispersion becomes perfectly 1D, where the DOS diverges as a power law with exponent $-1/2$.
    Our model produces a divergence with exponent $-0.205$, which is weaker than that of the ideal 1D case (iii). This can be understood by considering the case $a>0$, $c>0$, and $b=0$: the dispersion still hosts a band minimum at the origin but is extremely flat along the $y$-direction within a finite momentum window, giving $D(\Delta E)\propto(\Delta E)^{-1/4}$.

    Because $|b|\ll a,c$, the system behaves approximately as in the quadratic–quartic case, yielding an effective power-law DOS with an exponent slightly smaller than $-1/4$. However, since $b<0$ while $a>0$, the band also hosts a van Hove singularity that produces a logarithmic divergence at very low energies. The lower bound of the power-law regime in Fig.~\ref{Figmain}e reflects this crossover: for energies above the taco-band bandwidth ($\sim1$ meV), the log–log DOS plot shows a clear linear trend consistent with the power-law scaling, whereas for $\Delta E<10^{-3.5}$ eV ($\sim0.3$ meV) the curve deviates from linearity, signaling the reemergence of the underlying logarithmic divergence associated with the van Hove singularity.

    This analysis, purely based on energy bands, may significantly change at low electron fillings, where the Coulomb interaction generally dominates over the kinetic energy. The quasi-1D character is thus expected to emerge only at intermediate fillings, neither too small nor too large. Estimating the interaction scale as $U\approx14.4/(\epsilon d)$ eV, where $d=\sqrt{7}a_0/n\sim8.7/n$ \r{A} is the mean electron separation per moir\'e unit cell and $\epsilon = 6.24$ for WSe$_2$~\cite{hou2022quantification}, we obtain the filling-dependent interaction strength shown as a blue curve in Fig.~\ref{Figmain}f. When the Fermi energy is below about 1.5 meV, $U$ exceeds 40 times the Fermi energy, potentially leading to Wigner crystallization~\cite{wigner1934interaction,tanatar1989ground,drummond2009phase,monarkha2012two}, indicating a strongly correlated regime that can obscure the van Hove singularity. However, there exists an intermediate Fermi-energy window of the quasi-1D character before the interaction strength falls below the Wigner-crystal threshold. In this regime, the enhanced DOS and strong Fermi-surface nesting associated with the taco band favor band-structure-driven instabilities. Including realistic gate screening further suppresses the Coulomb energy [magenta curve in Fig.~\ref{Figmain}f], enlarging this window and making the underlying van Hove singularity and quasi-1D features more prominent.


\section*{Discussion}
Our work reveals that twisted bilayer WSe$_2$ at $\theta=21.8^\circ$ has an intriguing electronic structure termed {\it taco bands}, which uniquely combines three intriguing band properties: quasi-one-dimensional dispersion, concentric Fermi surfaces, and momentum-dependent spin polarization. We further demonstrate that this phenomenon is generic across twisted TMDs. Our analysis of three additional twisted bilayer TMDs shows that WS$_2$ likewise exhibits flat bands at 21.8$^\circ$, while MoSe$_2$ and MoS$_2$ have a different magic angle that is slightly larger than 21.8$^\circ$ (see details in SI Fig. S2).

Quasi-one-dimensional behavior usually arises from chain-like atomic bonding in real space, and manifests as approximately parallel Fermi surfaces extended across the two-dimensional BZ~\cite{giamarchi2004theoretical,wang2006new,watson2017multiband,kang2021band,du2023crossed,yu2023evidence,deng2025epitaxially,li2024imaging}. In such systems, low-energy excitations can show Tomonaga-Luttinger liquid (TLL) behavior in a finite temperature range~\cite{tomonaga1950remarks,luttinger1963exactly}. On the other hand, concentric Fermi surfaces emerge from Mexican-hat-type~\cite{wickramaratne2015electronic,seixas2016multiferroic,jiang2020van,jia2023tetragonal} or Rashba spin–orbit-coupled dispersions~\cite{berg2012electronic,hutchinson2016rashba}, where the local band minima form a ringlike shape such that Fermi surfaces are annular. Since their band structure is locally quasi-one-dimensional, correlation effects are strongly enhanced~\cite{castro2008low,ghazaryan2021unconventional}. However, TLL behavior is incompatible with an annular Fermi surface, where the radial (dispersive) and angular (dispersionless) directions are globally entangled around the ring.
The taco bands reconciles these limits: hexagonally concentric Fermi surfaces that contain extended quasi-one-dimensional segments. This geometry allows TLL-like responses to emerge on finite arcs while preserving a global concentric shape, enabling an intriguing interplay between 1D and 2D physics.
Moreover, each quasi-one-dimensional valley is strongly spin-polarized, resulting in alternating spin-up and spin-down texture in momentum space, similar to that in altermagnetic MnTe~\cite{vsmejkal2022emerging,lee2024broken,krempasky2024altermagnetic}.
In conclusion, the combination of pronounced one-dimensional character, concentric Fermi surfaces, and alternating spin texture in twisted WSe$_2$ provides a novel platform to probe interaction-driven instabilities.

\section*{References}
\putbib[references]

\section*{Data availability}
The data that support the plots within this paper and other findings of this study are available from the corresponding authors upon reasonable request. 

\section*{Acknowledgments}
This work was primarily supported by the NSF MRSEC DMR-2308817. We acknowledge the Texas Advanced Computing Center (TACC) at The University of Texas at Austin for providing computational resources that have contributed to the research results reported within this paper. This work also used computational resources from Stampede3 at The University of Texas at Austin through allocation DMR24024 and PHY250206 from the Advanced Cyberinfrastructure Coordination Ecosystem: Services and Support (ACCESS) program, which is supported by National Science Foundation grants 2138259, 2138286, 2138307, 2137603, and 2138296. 
S. H. L and J. A. were supported by a faculty start-up grant provided to J. A. by The University of Texas at Austin. S. H. L. was also supported by the Texas Quantum Institute Postdoctoral Fellowship.

\section*{Author contributions}
M. H. N. conceived the project. G. E., C. K. C. and M. H. N performed the first-principles calculations. S. H. L. and J. A. developed the low-energy continuum model. M. H. N. and J. A. supervised the project and wrote the first draft along with G. E. and S. H. L. Every author read and commented on the paper.

\section*{Competing interests}
The authors declare no competing interests.

\clearpage
\renewcommand{\baselinestretch}{1}

\section*{Figures}

\newpage

\clearpage
\newpage

\clearpage
\newpage



\clearpage
\newpage

\end{bibunit}


\clearpage

\appendix


\renewcommand{\appendixpagename}{\center\large Supplementary Information: ``\ourtitle''}

\appendixpage

\setcounter{page}{1}
\setcounter{section}{0}
\setcounter{figure}{0}
\setcounter{equation}{0}
\setcounter{table}{0}

\renewcommand{\thefigure}{S\arabic{figure}}
\renewcommand{\theequation}{S\arabic{equation}}
\renewcommand{\thesection}{S\arabic{section}}
\renewcommand{\thetable}{S\arabic{table}}

\author{
 Giovanny Espitia$^{1*}$,
 Seung Hun Lee$^{1*}$, Calvin Kaiyu Chiu$^{1}$, Junyeong Ahn$^{1\dagger}$ and Mit H. Naik$^{1\dagger}$
}



\maketitle 
\begin{affiliations}
 \item Department of Physics and Center for Complex Quantum Systems, The University of Texas at Austin, Austin, Texas, 78712, USA.
\end{affiliations}
\noindent 
*these authors contributed equally\\
$\dagger$ Corresponding authors:~\textcolor{blue}{{junyeong.ahn@austin.utexas.edu,~mit.naik@austin.utexas.edu}}

\begin{bibunit}
\section*{Methods}

\subsection{First-principles calculations}
Density functional theory calculations: 
Moir\'e structures were generated using the \texttt{Twister}\cite{naik2021twister} code. Structural relaxations were completed using classical force field calculations with \texttt{LAMMPS}~\cite{thompson2022lammps, plimpton1995lammps}, employing the Stillinger Weber~\cite{stillinger1985potential} potential for intralayer interactions and the Kolmogorov-Crespi potential~\cite{naik2019kolmogorov} for interlayer coupling. The relaxed configurations served as input for electronic structure calculations that were performed using the strictly localized atomic orbitals DFT framework implemented in the \texttt{SIESTA}\cite{soler2002siesta, garcia2020siestarecent} package. The generalized gradient approximation (GGA) in the Perdew, Burke, and Ernzerhof (PBE)\cite{perdew1996generalized} implementation was employed for the exchange-correlation functional. Fully relativistic norm-conserving pseudopotentials were used in the DFT calculations. For the DFT calculations, the real space grid to represent physical quantities was constrained to a cutoff of 160 Ry. For the structures twisted at $21.8^{\circ}$, the BZ was sampled using a 30x30x1 k - point grid without symmetry reduction. To sample the BZ of the other structures considered, the number of k - points was scaled in accordance to the moir\'e wavelength relative to the period at $21.8^{\circ}$. The DOS was subsequently computed and normalized per atom using a Gaussian smearing method according to the formula:
$$
g(E) = \frac{1}{N_{\text{atoms}}} \frac{1}{N_k} \sum_{n, \mathbf{k}} \frac{1}{\sigma \sqrt{\pi}} \exp\left(-\left(\frac{E - (E_{n, \mathbf{k}} - E_F)}{\sigma}\right)^2\right)
$$
where $N_{\text{atoms}}$ is the number of atoms in the supercell and  $N_k$ is the total number of k-points. The sum runs over all band indices $n$ and k-points $\mathbf{k}$. The term $E_{n, \mathbf{k}} - E_F$ represents the eigenvalues relative to the Fermi level, and $\sigma$ is the Gaussian broadening width, which was set to 0.009~eV. This yields a DOS with units of states per eV per atom. 

GW calculations: We computed the quasiparticle band gap correction for the twisted bilayer WSe$_2$ structures using single-shot $G_0W_0$ calculations \cite{hybertsen1986electron}, performed with the \texttt{BerkeleyGW} package \cite{deslippe2011berkeley}. These calculations were conducted on an AB-stacked unit cell whose interlayer spacing was set to match that of the twisted bilayers, thereby ensuring a similar screening environment. The dielectric matrix was constructed within the random-phase approximation (RPA) with a 30 Ry cutoff and 5109 unoccupied bands. We sampled the Brillouin-zone using a 12×12×1 k-mesh, accelerating convergence with the non-uniform neck sub-sampling method \cite{dajornada2017nonuniform}.

Spin texture: The spin texture of the $21.8^{\circ}$\,t--WSe$_2$ moiré superlattice was obtained from fully relativistic, non-collinear DFT calculations.  
At every band index $n$ and crystal momentum $\mathbf{k}$ we solved the Kohn–Sham equation for a two-component spinor
\[
\Psi_{n\mathbf{k}}(\mathbf{r}) \;=\;
\begin{pmatrix}
\psi_{n\mathbf{k}}^{\,\uparrow}(\mathbf{r}) \\[4pt]
\psi_{n\mathbf{k}}^{\,\downarrow}(\mathbf{r})
\end{pmatrix},
\]
whose upper and lower components give the probability amplitudes for finding the electron at position $\mathbf{r}$ with spin projection
$\uparrow$ or $\downarrow$ along the global $z$ axis, respectively.  
Each component is expanded in a plane-wave basis,
\[
\psi_{n\mathbf{k}}^{\,s}(\mathbf{r})
\;=\;
\sum_{\mathbf{G}}
c_{n\mathbf{k}}^{\,s}(\mathbf{G})\;
e^{i(\mathbf{k}+\mathbf{G})\cdot\mathbf{r}},
\qquad
s\in\{\uparrow,\downarrow\},
\]
where  $\mathbf{G}=h\,\mathbf{b}_1+k\,\mathbf{b}_2+\ell\,\mathbf{b}_3$ is a reciprocal-lattice vector of the moiré supercell ($h,k,\ell\in\mathbb{Z}$; $\mathbf{b}_i$ are its primitive reciprocal vectors), $c_{n\mathbf{k}}^{\,s}(\mathbf{G})\in\mathbb{C}$ is the plane-wave coefficient. The spin-resolved norm of a state is therefore
$\lVert\psi_{n\mathbf{k}}^{\,s}\rVert^{2}
=\sum_{\mathbf{G}}|c_{n\mathbf{k}}^{\,s}(\mathbf{G})|^{2}$,
and the expectation value of the Pauli operator $\sigma_z$ is
\[
\bigl\langle S_z \bigr\rangle_{n\mathbf{k}}
=\frac{\hbar}{2}\!
\left(
\lVert\psi_{n\mathbf{k}}^{\,\uparrow}\rVert^{2}
-
\lVert\psi_{n\mathbf{k}}^{\,\downarrow}\rVert^{2}
\right)
=\frac{\hbar}{2}\!
\left(
\sum_{\mathbf{G}}|c_{n\mathbf{k}}^{\,\uparrow}(\mathbf{G})|^{2}
-
\sum_{\mathbf{G}}|c_{n\mathbf{k}}^{\,\downarrow}(\mathbf{G})|^{2}
\right).
\]
To visualize this spin texture in the band structure plot, for each band, the color of each line segment (\textbf{k}-point) is determined by its corresponding $\langle S_z \rangle$ value. A custom colormap was used, mapping positive spin polarization (spin-up) to red and negative polarization (spin-down) to blue. States with a near-zero spin polarization, where the bands are effectively degenerate, are colored grey.

\begin{figure}[H]
\centering
 \includegraphics[scale=0.75]{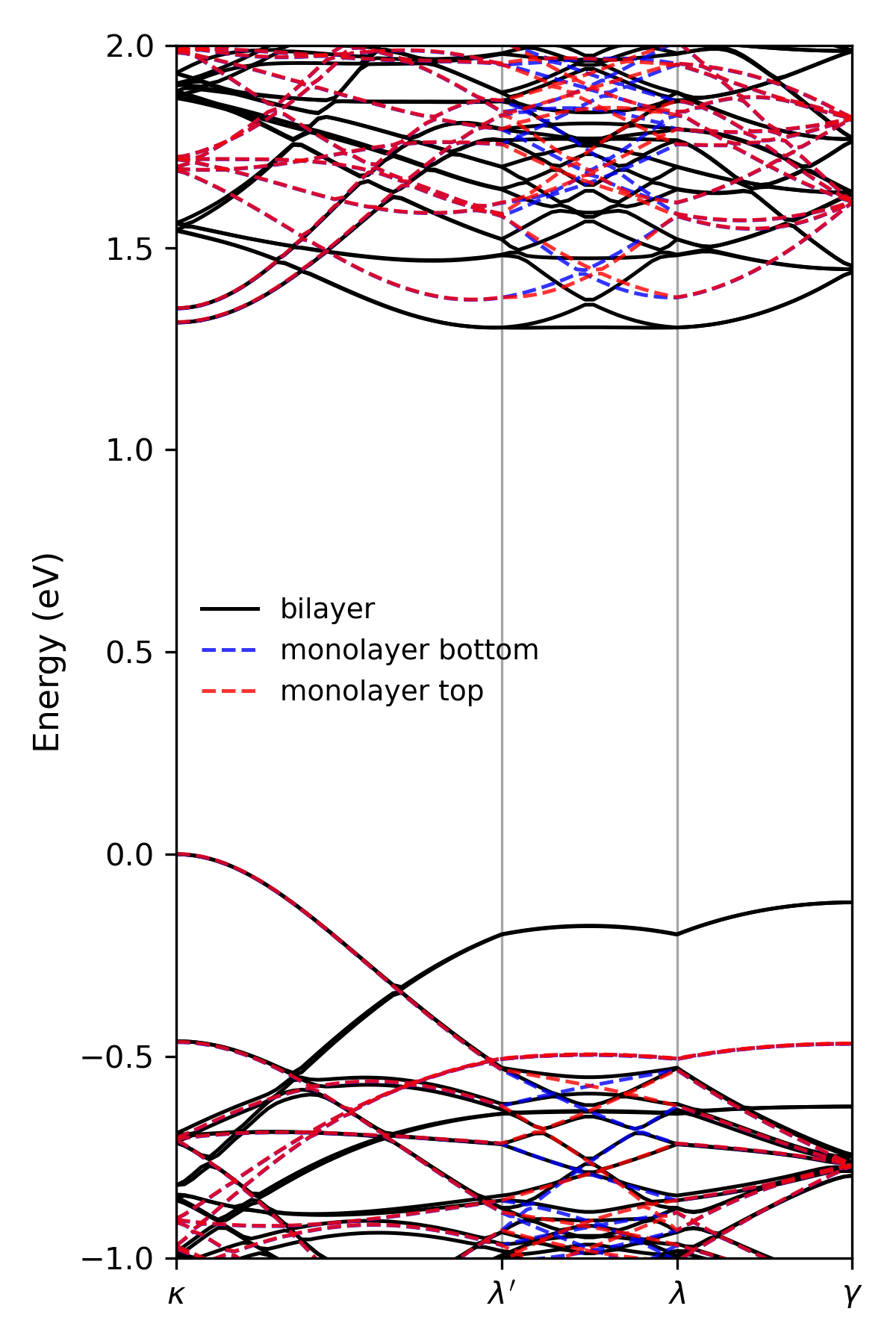}
    \caption{\textbf{Band structures of monolayer and twisted bilayer WSe$_2$.} The plot shows the density functional theory band structure for twisted WSe$_2$ at a 21.8$^\circ$ twist angle alongside the individual monolayer band structures. The conduction band edges of both isolated monolayers are higher in energy compared to the bilayer bands due to the absence of interlayer hybridization. 
}
    \label{fig:s1}
\end{figure}

\begin{figure}[H]
 \includegraphics[width=14.5cm]{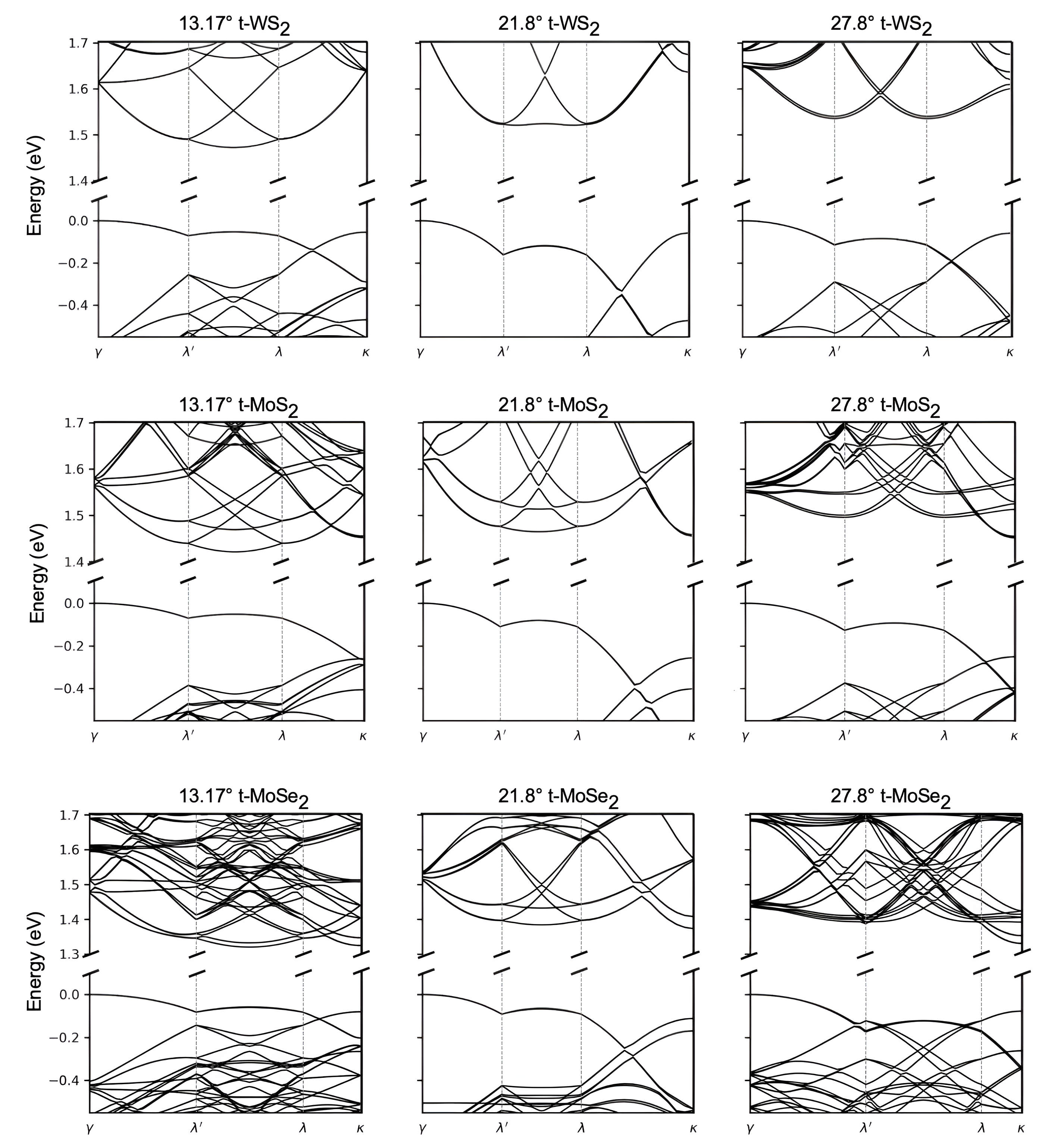}
    \caption{\textbf{Twist angle dependence of electronic band structure for various transition metal dichalcogenides.} The plot presents the band structures of t-WS$_2$, t-MoS$_2$, and t-MoSe$_2$ at 13.17$^{\circ}$, 21.8$^{\circ}$, and 27.8$^{\circ}$. t-WS$_2$ exhibits a flat band between the $\lambda$–$\lambda^{\prime}$ valleys at the 21.8$^{\circ}$ twist angle, consistent with observations in t-WSe$_2$. In contrast, t-MoS$_2$ and t-MoSe$_2$ display a more dispersive electronic structure at 21.8$^{\circ}$.} 
    \label{fig:s2}
\end{figure}

    \subsection{Moir\'e band model}
    In this section, we construct a continuum model that describes the band flattening along the $\lambda-\lambda'$ direction in twisted bilayer WSe$_2$ around a specific commensurate twist angle of $21.8^{\circ}$, where $\lambda$ ($\lambda'$) denotes the band minimum of the $\gamma-\kappa$ ($\gamma-\kappa'$) line. Here, $\gamma=(0,0)$, $\kappa=a_0^{-1}\sin(\theta/2)(-4\sqrt3\pi/3,4\pi/3)$ and $\kappa'=a_0^{-1}\sin(\theta/2)(-4\sqrt3\pi/3,-4\pi/3)$ represent high-symmetry points of the moir\'e Brillouin zone (mBZ). $a_0$ is the lattice constant of monolayer WSe$_2$. We use lowercase Greek letters to denote points in the mBZ to distinguish them from those of the top and bottom monolayers.

    \textit{Monolayer WSe$_2$ tight-binding model}---
    We adopted the Hamiltonian and parameters from Ref.~\cite{liu2013three} for the tungsten $d$-orbital tight-binding model without spin-orbit coupling in the basis of ($d_{z^2},d_{xy},d_{x^2-y^2}$).
    
    \begin{figure}[H]
        \centering
  	\includegraphics[width=0.6\linewidth]{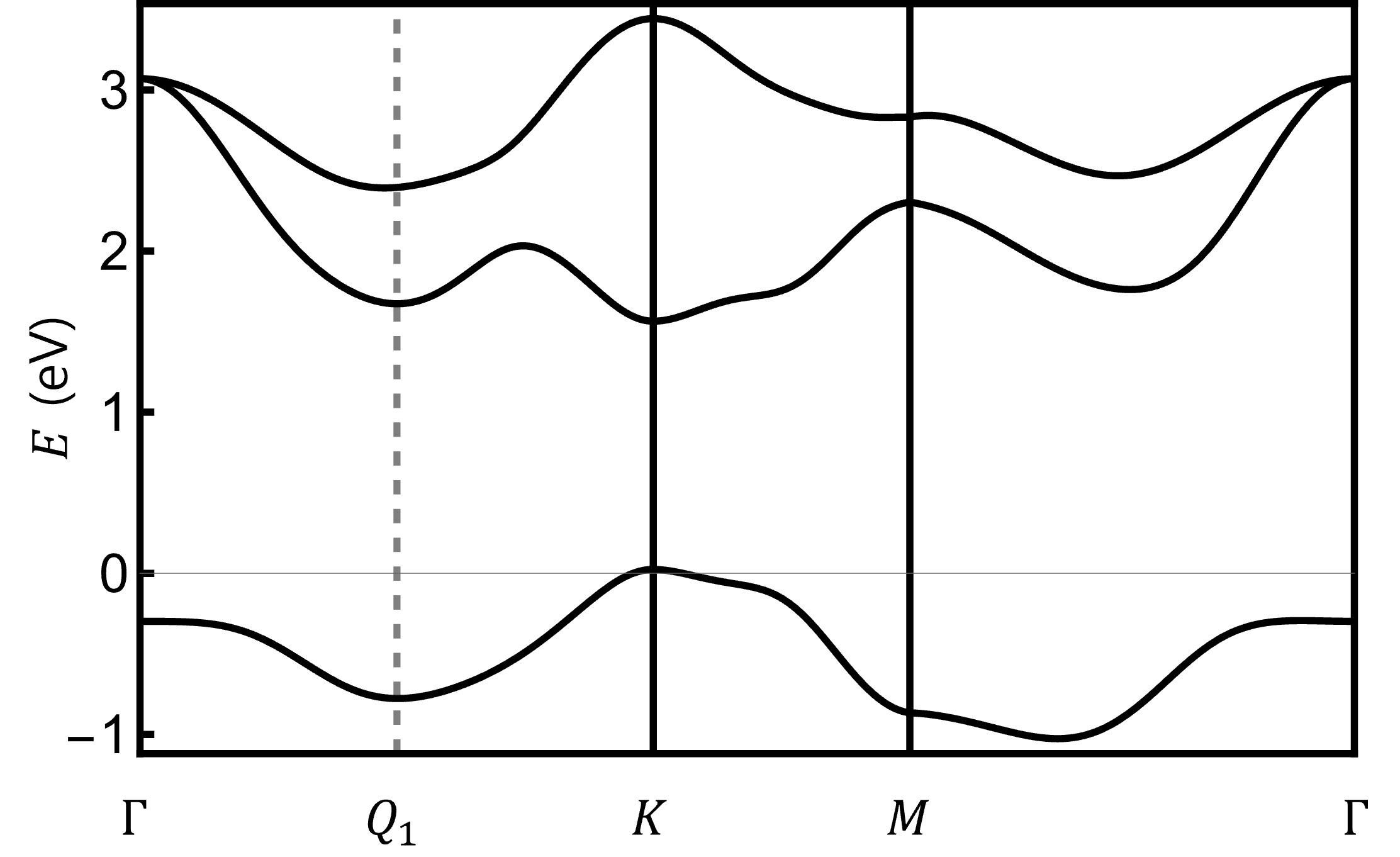}
  	\caption{
        {\bf {Band structure obtained from the monolayer WSe$_2$ tight-binding model.}}
            $Q_1=a_0^{-1}(2\pi/3,0)\approx a_0^{-1}(2.0944,0)$ is the midpoint between $\Gamma$ and $K$. The dispersion of the second band along the $\Gamma-K$ line has a minimum at $\Lambda_1=a_0^{-1}(2.0975,0)$, very close to $Q_1$.
  	}\label{mono}
    \end{figure}

    Fig.~\ref{mono} shows the band structure obtained from this tight-binding model.
    Without doping, only the bottom band is occupied.
    Our main interest is on the minima of the lowest unoccupied band located close to the midpoints of the $\Gamma-K$ and $\Gamma-K'$ lines.
    We refer to these band minima as the $\Lambda$ valleys, and we denote the corresponding midpoint momenta as the $Q$ points. Within the first Brillouin zone (BZ), there are six inequivalent $Q$ points. We denote them as $Q_i$ with $i=1,2,\cdots,6$, where
    \begin{equation}
        Q_1=a_0^{-1}(2\pi/3,0).
    \end{equation}
    The actual minimum on the $\Gamma-K$ line, $\Lambda_1=a_0^{-1}(2.0975,0)$, is very close to $Q_1\approx a_0^{-1}(2.0944,0)$. The other $Q$ points are generated by successive 60$^\circ$ rotations, i.e.
    \begin{equation}
        Q_i=[R(\pi/3)]^{(i-1)}Q_1,
    \end{equation}
    where $R(\phi)=\begin{pmatrix}
        \cos\phi & -\sin\phi \\
        \sin\phi & \cos\phi
    \end{pmatrix}$ is a rotation matrix.
    The energy dispersion near $Q_i$ is quadratic. The effective mass of the $\Lambda_1$ valley is
    \begin{align}
        m_x&\approx0.849/(a_0^2~\textrm{eV}),\\
        m_y&\approx0.749/(a_0^2~\textrm{eV}),
    \end{align}
    in the $k_x$ and $k_y$ directions, respectively.
    
\begin{figure}[H]
        \centering
  	\includegraphics[width=0.8\linewidth]{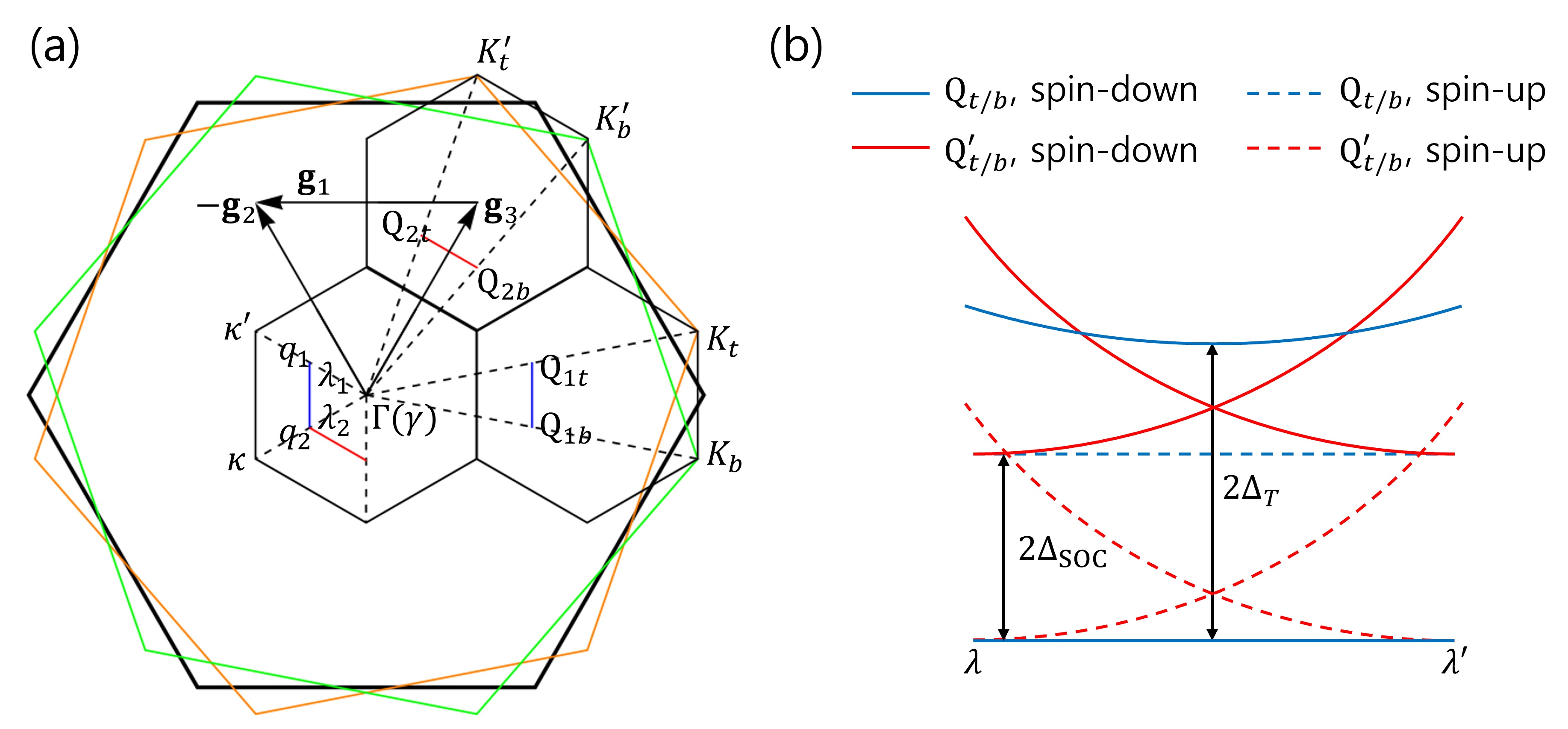}
  	\caption{
        {\bf {Schematic description of the $\lambda-\lambda'$ valley continuum model.}}
  		(\textbf{a}) The large black, green, and orange hexagons represent the BZs of the untwisted, top, and bottom WSe$_2$ layers with the twist angle $\theta=21.8^\circ$, respectively. The smaller hexagons indicate the mBZs of the twisted bilayer WSe$_2$. $\textbf g_1$, $\textbf g_2$ and $\textbf g_3=-\textbf g_1-\textbf g_2$ denote the reciprocal lattice vectors of the mBZ. The band dispersions near the $\lambda_1$ ($\lambda_2$) point are obtained by the translation of dispersions around the $Q_{1t}$ and $Q_{6b}$ ($Q_{1b}$ and $Q_{2t}$) points by the reciprocal lattice vectors.
  		(\textbf{b}) A simplified schematic of the low-energy unoccupied band structure. Note that the effective masses of the red and blue bands differ due to the anisotropic dispersions near the $\lambda(=\lambda_2)$ and $\lambda'(=\lambda_1)$ valleys.
  	}\label{Fig1}
    \end{figure}

    After twisting, the $Q_1$ point of the bottom and top layers, denoted as $Q_{1b}$ and $Q_{1t}$, are folded to the $q_1$ and $q_2$ points which are the midpoints of the $\gamma-\kappa$ and $\gamma-\kappa'$ lines in the mBZ, as shown in Fig.~\ref{Fig1}a. Here, 
    \begin{equation}
        q_1=a_0^{-1}2\pi\sin\frac{\theta}{2}(\frac{1}{\sqrt {3}},\frac{1}{3}),
    \end{equation}
    and
    \begin{equation}
        q_i=[R(\pi/3)]^{(i-1)}q_1.
    \end{equation}
    While we are interested in the region close to these points, the actual band minima $\lambda_i$s of twisted bilayer WSe$_2$ at the twist angle $21.8^\circ$ are not located at $q_i$s. As our first-principles calculations show, the distance between $\lambda_i$ and $q_i$ is greatly amplified by twisting, compared to the original small deviation between $\Lambda_i$ and $Q_i$.
    Although both $\lambda_i$ and $q_i$ lie along the $\gamma-\kappa$ or $\gamma-\kappa'$ line, the distance from $\gamma$ to $\lambda_i$ is about 0.7 times that from $\gamma$ to $q_i$. This significant shift is due to the moir\'e potential that applies to each valley. We discuss this effect in detail later in the section \textit{moir\'e potential}.
   
    Upon including spin–orbit coupling in the model, the band near the $\Lambda$ valleys splits into two almost spin-polarized branches separated by approximately $0.24$ eV~\cite{liu2018nature}. Importantly, between these two spin branches, the spin-up band has lower energy at the $\Lambda_i$ valleys with odd $i$, whereas the spin-down band lies lower at the $\Lambda_i$ valleys with even $i$, reflecting the alternating sign of the spin–orbit coupling along the $\Gamma-K$ versus $\Gamma-K'$ directions.

    In the following, we construct a spin-polarized continuum model of twisted WSe$_2$, based on the quadratic kinetic-energy dispersion $(k_x-[\textbf k_{Q_i}]_x)^2/(2m_x)+(k_y-[\textbf k_{Q_i}]_y)^2/(2m_y)$ at the $Q_i$ valleys of monolayer WSe$_2$.
    We show that the partially flat band—dispersionless along one direction but dispersive along the other—which we refer to as the \textit{taco} band, arises from interlayer tunneling that opens a gap between bands from two nearby valleys.
    Since the valley momenta lie on the $\gamma-\kappa$ or $\gamma-\kappa'$ lines, the band flattening appears along the lines connecting these valleys. Consequently, the taco band emerges along the edges of the hexagon formed by the six $\lambda_i$ points.
    
\begin{figure}[H]
        \centering
  	\includegraphics[width=0.6\linewidth]{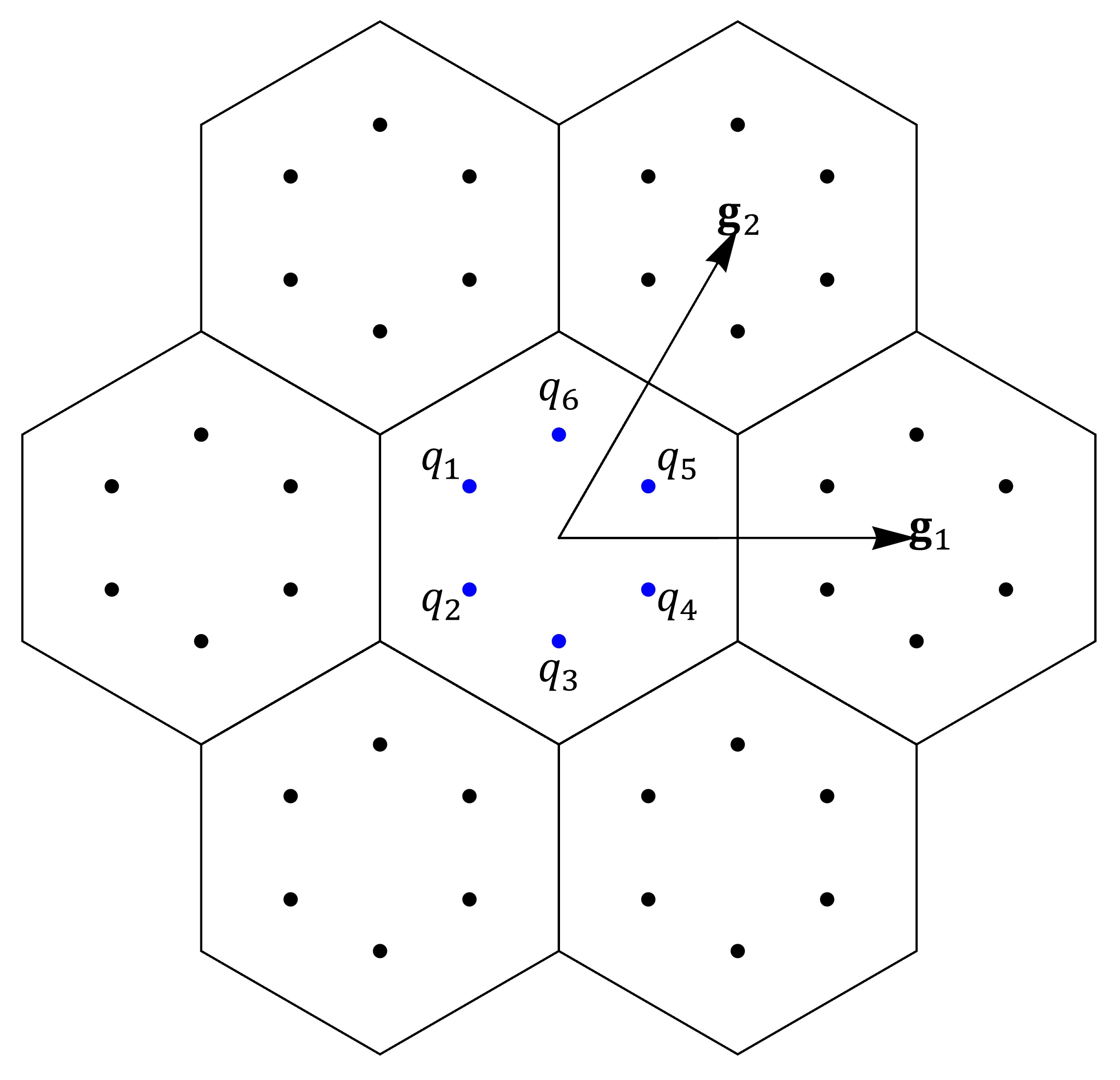}
  	\caption{
        {\bf {Reciprocal lattice of valleys in momentum space.}}
  		The blue dots represent the $q_i$ valleys in the first mBZ. The black dots represent their replicas obtained by successive translations with the moir\'e reciprocal lattice vectors $\textbf g_1$ and $\textbf g_2$.
  	}\label{valley}
    \end{figure}

    \textit{Continuum model Hamiltonian of twisted WSe$_2$}---
    The Hamiltonian operator is in the form of
    \begin{equation}
    \hat H=\sum_{\textbf k,\textbf k'}(\psi_{t,\textbf k}^\dagger,\psi_{b,\textbf k'}^\dagger)H(\textbf k,\textbf k')(\psi_{t,\textbf k},\psi_{b,\textbf k'})^T,
    \end{equation}
    where $\psi_{l,n,\textbf k}^\dagger$ ($\psi_{l,n,\textbf k}$) is the creation (annihilation) operator of an electron in a Bloch state with the layer index $l=t,b$, valley index $n$, and momentum $\textbf k$. The indices with and without a prime sign denote that the Bloch state is from the bottom and top layer, respectively.
    In momentum space, the valleys form a reciprocal lattice with periodicity
    \begin{equation}
        \textbf g_i=[R(+\theta/2)-R(-\theta/2)]\textbf G_i
    \end{equation}
    for $i=1,2,3$, where $\textbf G_1=a_0^{-1}(0,4\pi/\sqrt 3)$ and $\textbf G_i=[R(2\pi/3)]^{(i-1)}\textbf G_1$. Thus, as in Fig.~\ref{valley}, the valley center in momentum space satisfies $\textbf k_n=\textbf k_{q_i}+z_1(n)\textbf g_1+z_2(n)\textbf g_2$ where $z_1(n)$ and $z_2(n)$ are integers.

    In the Bloch state basis, diagonal elements of $H(\textbf k,\textbf k')$ take the form
    \begin{align}
        H_{t,nn}(\textbf k)&\equiv [H(\textbf k,\textbf k')]_{t,n;t,n}=\langle t,n, \textbf k|\hat H|t,n, \textbf k\rangle=\frac{(\textbf k-\textbf k_n)^2}{2m}+\Delta_n(\textbf k),\\
        H_{b,n'n'}(\textbf k')&\equiv[H(\textbf k,\textbf k')]_{b,n';b,n'}=\langle b, n', \textbf k'|\hat H|b,n', \textbf k'\rangle=\frac{(\textbf k'-\textbf k'_n)^2}{2m}+\Delta_{n'}(\textbf k'),
    \end{align}
    where $|l,n,\textbf k\rangle=\psi_{l,n,\textbf k}^\dagger|0\rangle$, $m$ is the effective mass of an electron near the valleys, and $\Delta_n(\textbf k)$ represents the moir\'e potential. Meanwhile, the off-diagonal elements of $[H(\textbf k,\textbf k')]_{t,n;b,n'}$ describe the interlayer tunneling $\Delta_T(n,\textbf k;n',\textbf k')$ between two Bloch states $|t,n,\textbf k\rangle$ and $|b,n',\textbf k'\rangle$.

    \textit{Moir\'e potential}---
    $\Delta_{n}(\textbf k)$ is a phenomenological term that incorporates all effects of the other layer, except for the direct tunneling between two valleys, which is included in $\Delta_{T}(n,\textbf k;n',\textbf k')$.
    $\Delta_{n}(\textbf k)$ is periodic in the mBZ and can therefore be expanded in a Fourier series of moiré reciprocal lattice vectors. For simplicity, we retain the leading harmonic term, which takes the form $\cos{(\textbf a \cdot \textbf k+\varphi)}$, where $\textbf a$ is a moir\'e lattice vector and $\varphi$ is a constant phase.
    We note that non-zero $\varphi$ is allowed in $\Delta_{n}(\textbf k)$ even though the full system has time-reversal symmetry, because we are currently looking at only one spin sector of the Hamiltonian. In the full Hamiltonian, $\Delta_{q_i}(\textbf k)$ is symmetry-related to another $\Delta_{q_j}(\textbf k)$ with opposite spin by time-reversal, but $\Delta_n(\textbf k)$ itself is not required to be an even function of $\textbf k$.
    Imposing the three-fold rotational symmetry around the $z$-axis ($C_{3z}$), we have
    \begin{equation}
        \Delta_{n}=A\cos{(\textbf a_n\cdot\textbf k+\varphi)},
    \end{equation}
    where $A$ is the amplitude of the moir\'e potential and $\textbf a_n$ is a moir\'e lattice vector that is parallel to $\textbf k_n$ valleys inside the first mBZ. For example, for the $q_1$ and $q_2$ valleys in Fig.~\ref{Fig1}, the corresponding moir\'e lattice vectors are $\textbf a_{q_1,q_2}=a_0(1/\sin{(\theta/2)})(-\sqrt{3}/4,\pm1/4)$.
    Fitting the DFT data, we obtain
    \begin{align}
        A&=-0.12115~\textrm{eV},\\
        \varphi&=0.5.
    \end{align}

    \begin{figure}[H]
        \centering
  	\includegraphics[width=0.6\linewidth]{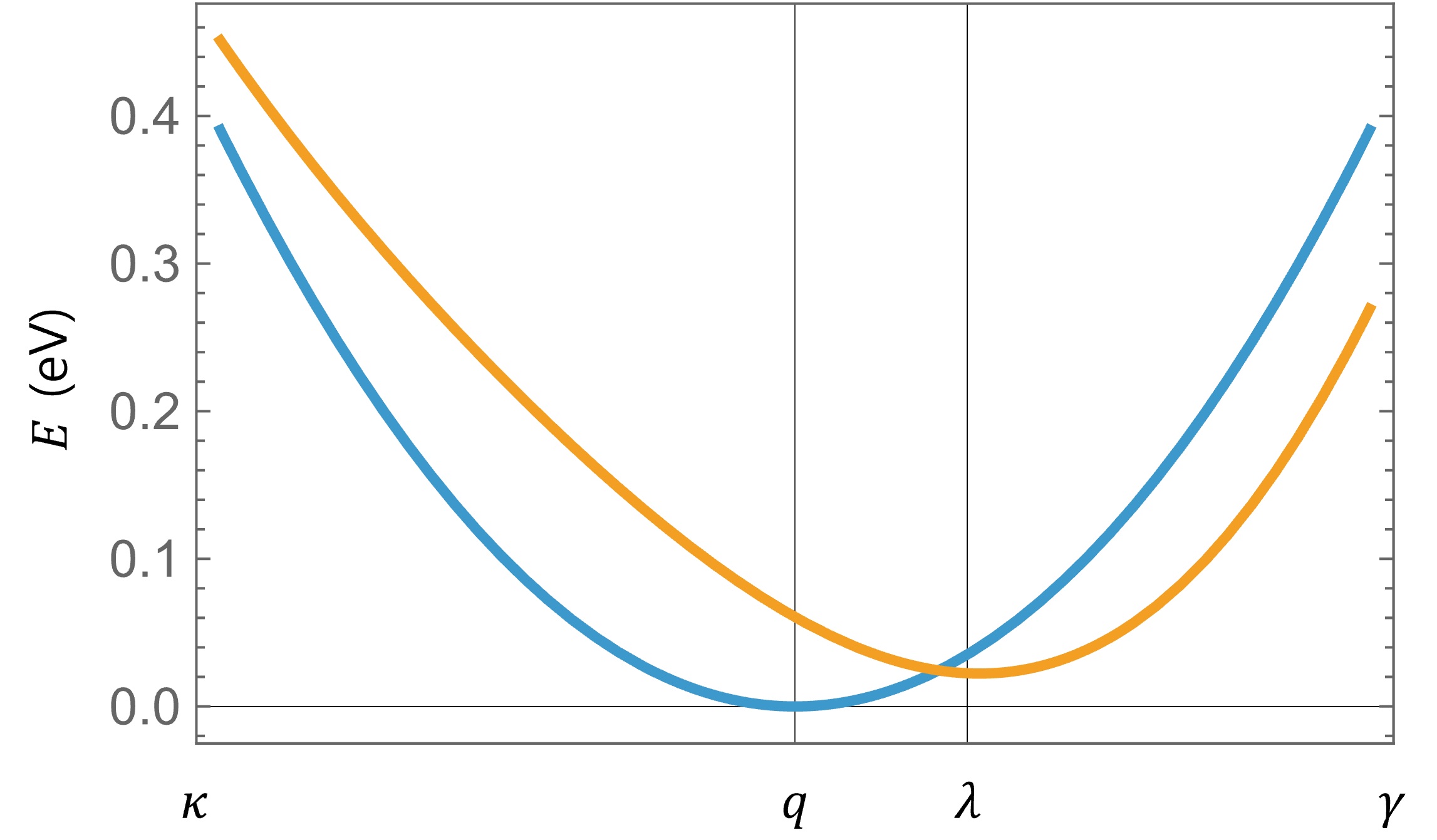}
  	\caption{
        {\bf {Shift of the band minimum along $\kappa-\gamma$ due to the moir\'e potential.}}
            The blue and orange lines show the energy dispersion of the continuum model (without interlayer tunneling) near the $q_i$ valley along the $\kappa-\gamma$ line, for $A=0$ and $(A=-0.12115\text{ eV},\varphi=0.5)$, respectively.
  	}\label{shift}
    \end{figure}
    
    As shown in Fig.~\ref{shift}, the effect of the moiré potential in our model is to shift the band minima from the $q_i$ valleys to the $\lambda_i$ valleys. The parameters $A$ and $\varphi$ are determined by fitting to the first-principles results at $21.8^\circ$.

    \textit{Interlayer tunneling}---
    $\Delta_T(n,\textbf k;n',\textbf k')$ is obtained using the two-center approximation~\cite{bistritzer2011moire}. The hopping integral between two Bloch states $|n,\textbf k\rangle$ and $|n',\textbf k'\rangle$ is given by
    \begin{align}\label{tunneling}
        \Delta_T(n,\textbf k;n',\textbf k')&\equiv\langle t, n,\textbf k|\hat H|b,n',\textbf k'\rangle\\\nonumber
        &=\sum_{\textbf R,\textbf R'}\frac{e^{i\textbf k'\cdot\textbf R'-i\textbf k\cdot\textbf R}}{\sqrt{NN'}}\langle t,n,\textbf R|\hat H|b,n',\textbf R'\rangle\\
        \nonumber
        &=\sum_{\textbf R,\textbf R'}\frac{e^{i\textbf k'\cdot\textbf R'-i\textbf k\cdot\textbf R}}{\sqrt{NN'}}t_{nn'}(\textbf R'-\textbf R)\\\nonumber
        &=\sum_{\textbf R,\textbf R'}\frac{e^{i\textbf k'\cdot\textbf R'-i\textbf k\cdot\textbf R}}{NN'}\sum_{\textbf q}e^{-i\textbf q\cdot(\textbf R'-\textbf R)}t_{nn'}(\textbf q)\\\nonumber
        &=\sum_{\textbf q}t_{nn'}(\textbf q)\sum_{\textbf R,\textbf R'}\frac{e^{i(\textbf k'-\textbf q)\cdot\textbf R'-i(\textbf k-\textbf q)\cdot\textbf R}}{NN'}\\\nonumber
        &=\sum_{\textbf q}t_{nn'}(\textbf q)\sum_{\textbf G,\textbf G'}\delta_{\textbf k-\textbf q,\textbf G}\delta_{\textbf k'-\textbf q,\textbf G'}e^{-i\textbf G\cdot\textbf d}\\\nonumber
        &=\sum_{\textbf G,\textbf G'}\delta_{\textbf k+\textbf G,\textbf k'+\textbf G'}t_{nn'}(\textbf k+\textbf G)e^{i\textbf G\cdot\textbf d},
    \end{align}
    where $N$ and $N'$ are the numbers of unit cells in the respective layers, and $|n,\textbf R\rangle$ and $|n',\textbf R'\rangle$ represent the Wannier states localized at $\textbf R=j_1\textbf a_1+j_2\textbf a_2-\textbf d$ and $\textbf R'=j'_1\textbf a'_1+j'_2\textbf a'_2$, respectively. For $i=1,2$, $j_i$ and $j'_i$ are arbitrary integers, and $\textbf a_i$ and $\textbf a'_i$ are basis lattice vectors of the corresponding layers. $\textbf d$ denotes the lateral displacement between the two layers.

    Equation~\eqref{tunneling} implies that for a given $\textbf k$, only $\textbf k'$ that satisfy $\textbf k+\textbf G=\textbf k'+\textbf G'$ yields non-zero value for the interlayer tunneling. Thus, for a Bloch state from the $Q_{1b}$ valley, 
    Assuming that $t_{nn'}(\textbf R'-\textbf R)$ decays exponentially with the distance $|\textbf R'-\textbf R|$, its Fourier transform $t_{nn'}(\textbf q)$ also decays exponentially with $|\textbf q|$. Therefore, we consider only a few $\langle t,n,\textbf k|\hat H|b,n',\textbf k'\rangle$ with  smallest values of $\textbf k+\textbf G$, while truncating out larger $\textbf G$ cases. In particular, at a momentum near the $Q_i$ ($Q'_i$) valleys, $\textbf G=0$ is the only one for which $\textbf k+\textbf G$ lies within the first BZ. Hence, $\textbf G=0$ yields the dominant contribution, and we obtain $\Delta_T(n,\textbf k;n',\textbf k')=t_{nn'}(\textbf k)$ for $\textbf k'$s such that $\textbf k'=\textbf k-\textbf G'$. Since the Bloch states with $\textbf k'$s for large $\textbf G'$s have very high energies when they are folded back into the first mBZ, we only consider $\textbf G=\textbf G'=0$ cases in our model.
    For example, from the lowest unoccupied states (which are the spin-up states) at $Q_{1b}$ and $Q_{1t}$ valleys in rotated monolayer WSe$_2$ BZs, we obtain the following Hamiltonian for the $q_1$ and $q_2$ valleys in mBZ:
    \begin{align}\label{H12}
        &H_\uparrow(\textbf k)=\nonumber\\
        &\begin{pmatrix}
            \sum_{\alpha=x,y}\frac{[R(-\theta/2)(\textbf k-\textbf k_{q_1})]_\alpha^2}{2m_\alpha}{2m_y}+A\cos{(\textbf a_{q_1}\cdot\textbf k+\varphi)} & t_{12}(\textbf k) \\
            t_{12}^*(\textbf k) & \sum_{\alpha=x,y}\frac{[R(\theta/2)(\textbf k-\textbf k_{q_2})]_\alpha^2}{2m_\alpha}+A\cos{(\textbf a_{q_2}\cdot\textbf k+\varphi)}
        \end{pmatrix}.
    \end{align}

    \subsection{Low-energy band structure along $\lambda-\lambda'$}
    In this section, we provide a detailed analysis of the low-energy band structure along the $\lambda-\lambda'$ direction, complementing the results discussed in the section \textbf{Magic angle near 21.8$^{\circ}$ in the low-energy model} of the main text. 

    \textit{Model Hamiltonian}---
    Along the $\lambda_1-\lambda_2$ line where $k_x$ is fixed, the Hamiltonian in Eq.~\eqref{H12} can be expressed as a function of $k_y$ in a simpler form:
    \begin{equation}\label{effH}
        H_\uparrow(k_y)=
        \begin{pmatrix}
            \frac{(k_y-k_{0y})^2}{2m} & t(k_y) \\
            t^*(k_y) & \frac{(k_y+k_{0y})^2}{2m}
        \end{pmatrix},
    \end{equation}
    where $k_{0y}\approx0.700\times(2\pi/3)\sin(\theta/2)$ $a_0^{-1}$ with $a_0$ denoting the lattice constant of monolayer WSe$_2$.
    $m\approx0.80177~a_0^{-2}$ eV$^{-1}$ is the effective mass along the $\lambda_1-\lambda_2$ direction, obtained by taking the inverse of second derivative of diagonal elements in Eq.~\eqref{H12}.

    \textit{Symmetry constraints of the effective Hamiltonian}---
    Twisted bilayer WSe$_2$ has time-reversal $T$ symmetry, three-fold rotation $C_{3z}$ symmetry around the $z$-axis, and two-fold rotation symmetries around three different in-plane axes which are related to each other by $C_{3z}$. In our calculation, one of the in-plane axis is taken to be the $y$-axis, so we denote the twofold rotation as $C_{2y}$, $C_{3z}C_{2y}C_{3z}^{-1}$, and $C_{3z}^2 C_{2y}C_{3z}^{-2}$.
    The effective Hamiltonian in Eq.~\eqref{effH} , which involves only two nearby valleys $\lambda$ and $\lambda'$ with the same spin, is not invariant under $T$, $C_{3z}$ and $C_{2y}$ individually.
    However, the line connecting $\lambda$ and $\lambda'$ is invariant under $C_{2y}T$, the combination of $T$ and $C_{2y}$. The representation of $C_{2y}T$ is given by
    $\begin{pmatrix}
        0 & 1 \\
        1 & 0
    \end{pmatrix}\mathcal{K}$ where $\mathcal{K}$ is complex conjugation.
    Since $H(k_y)$ satisfy $(C_{2y}T)H(k_y)(C_{2y}T)^{-1}=H(-k_y)$, we find that $t(k_y)=t(-k_y)$. Thus, in the lowest orders of $k_y$, 
    \begin{equation}
        t(k_y)\approx \tilde w_0+\tilde w_2 k_y^2,
    \end{equation}
    where both $\tilde w_0$ and $\tilde w_2$ are complex numbers.

    \begin{figure}[H]
        \centering
  	\includegraphics[width=1.0\linewidth]{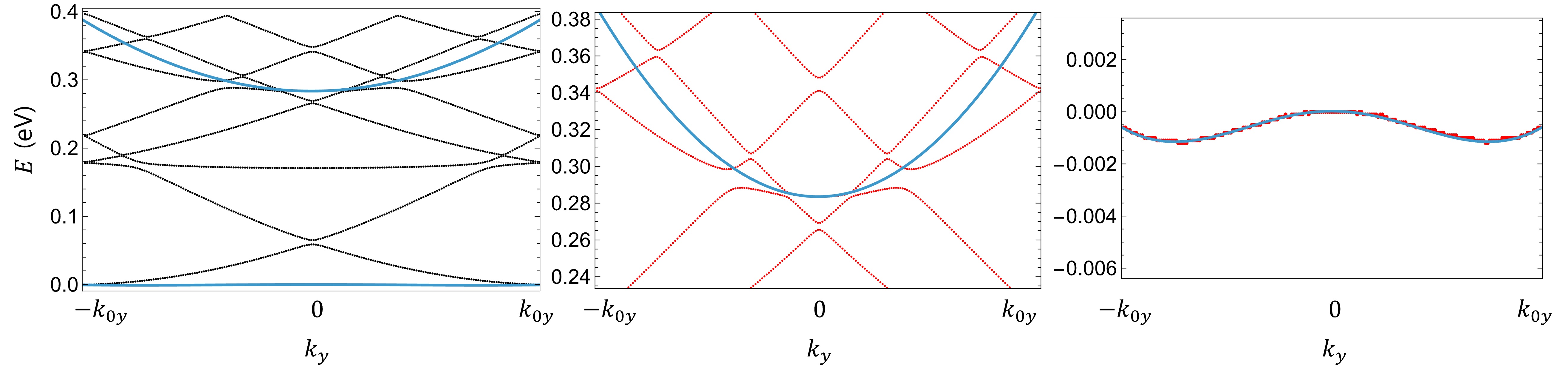}
  	\caption{
        {\bf {Numerical calculation results and fitted data.}}
  		The blue solid lines represent the energy dispersions of the effective Hamiltonian in Eq.~\ref{effH}. The left panel shows the agreement between the numerically calculated band structure and the model dispersion over a wide energy range. The middle and right panels provide a closer view, highlighting the agreement of $E_+(k_y)$ and $E_-(k_y)$ with the numerical data, respectively.
  	}\label{Fig2}
    \end{figure}
    
    \textit{Fitted parameters}---
    The eigenvalues of Eq.~\eqref{effH} are given by
    \begin{equation}\label{epm}
        E_\pm(k_y)=\frac{k_{0y}^2 m+k_y^2 m\pm2\sqrt{k_{0y}^2 k_y^2 m^2+m^4|t(k_y)|^2}}{2m^2}+E_0,
    \end{equation}
    where $E_0$ is an additional constant energy shift introduced to match the conduction band energy level obtained from first-principles calculations at $k_y=0$.
    We rewrite $\tilde w_0=e^{i\phi_o}w_0$ and $\tilde w_2=e^{i\phi_o}e^{i\phi_r}w_2$, in which $\phi_o$ and $\phi_r$ represent the overall and relative phases of $\tilde w_0$ and $\tilde w_2$, respectively. Then $|t(k_y)|^2=w_0^2+2w_0 w_2 k_y^2\cos{\phi_r}+(w_2 k_y^2)^2$. Since only the relative phase $\phi_r$ enters the expression for $E_\pm(k_y)$, the overall phase $\phi_o$ does not affect the band dispersion and cannot be determined from it.
    We determine the model parameters by fitting the energy dispersion to the results from first-principles calculations. The value of $w_0$ can be directly extracted from the relation $E_+(0)-E_-(0)=2w_0$, giving
    \begin{align}
        w_0\approx0.14175\text{ eV}.
    \end{align}
    Fig.~\ref{Fig2} shows the fitting results. With fitted parameters
    \begin{align}
        w_2&\approx0.49357~a_0^2\text{ eV},\\ 
        \phi_r&\approx1.10536,
    \end{align}
    the model almost perfectly reproduces the dispersion of the lower (flat) band and accurately captures the upper band’s behavior near $k_y=0$.

    \textit{Bandwidth and magic angle}---
    Using the same parameter values for $m$, $w_0$, $w_2$, and $\phi_r$, we compute the band dispersion for different twist angles $\theta$. As expected, for a small twist angle (e.g., $\theta=18^\circ$), the dispersion of the lower band is convex downward, indicating that the two valleys are close in momentum space and the interlayer tunneling dominates, pushing bands downward at the midpoint of $\lambda$ and $\lambda'$.
    In contrast, for a large twist angle (e.g., $\theta = 25^\circ$), the dispersion becomes convex upward. This is because the two valleys are far apart, and the kinetic energy of electrons within each valley becomes the dominant contribution.
    At a specific intermediate twist angle, however, these two competing effects balance each other, resulting in an extremely flat band.
    
    \begin{figure}[H]
        \centering
  	\includegraphics[width=1.0\linewidth]{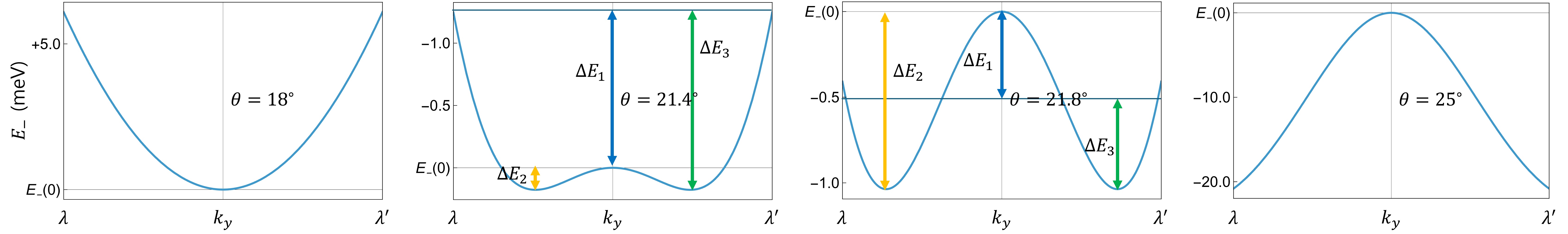}
  	\caption{
        {\bf {Variation of the dispersion of $E_-(k_y)$ with respect to $\theta$.}}
  		In each panel, the $y$-axis is referenced to $E_-(k_y = 0)$ for the corresponding value of $\theta$.
  	}\label{Fig5}
    \end{figure}
    
    As illustrated in the main text, the bandwidth along the $\lambda-\lambda'$ line is given by the largest among the three energy differences:
    \begin{align}
        \Delta E_1&=|E_-(0)-E_-(k_{0y})|,\\ \Delta E_2&=|E_-(0)-E_{\textrm{min}}|,\\
        \Delta E_3&=|E_-(k_{0y})-E_{\textrm{min}}|,
    \end{align}
    where $E_{\textrm{min}}$ denotes the minimum---if present---of $E_-(k_y)$ between $k_y=0$ and $k_y=k_{0y}$ (See Fig.~\ref{Fig5}).
    The minima appear at
    \begin{align}
        k_y=\pm&\frac{1}{|mw_2|\sqrt{2(1-4m^2 w_2^2)}}\bigg[(1-4m^2 w_2^2)(k_{0y}^2+2m^2 w_0 w_2\cos\phi_r)\nonumber\\
        &-\sqrt{(1-4m^2 w_2^2)(k_{0y}^4-2m^4 w_0^2 w_2^2+4k_{0y}^2 m^2 w_0 w_2\cos\phi_r+2m^4 w_0^2 w_2^2\cos{2\phi_r})}\bigg]^{1/2}.
    \end{align}
    The explicit form $E_{\textrm{min}}$ is given by
    \begin{align}
        E_{\textrm{min}}=&\frac{1}{2m w_2^2(1-4m^2 w_2^2)}\bigg[k_{0y}^2(1-6m^2 w_2^2+8m^4 w_2^4)+2m^2 w_0 w_2(1-4m^2 w_2^2)\cos\phi_r\nonumber\\
        &-(1-4m^2 w_2^2)\sqrt{(1-4m^2 w_2^2)(k_{0y}^4-2m^4 w_0^2 w_2^2+4k_{0y}^2 m^2 w_0 w_2\cos\phi_r+2m^4 w_0^2 w_2^2\cos{2\phi_r})}\bigg].
    \end{align}
    We note that $\Delta E_1=\Delta E_3-\Delta E_2$ serves as a good indicator for the “magic angle”. This is because the bandwidth, given by $\max{(\Delta E_1,\Delta E_2,\Delta E_3)}$, reaches its minimum when $\Delta E_2=\Delta E_3$. The magic angle, defined as the angle at which $\Delta E_1=0$, is given by
    \begin{equation}
        \theta_c=2\arcsin{\Bigg[\frac{3}{\pi}\frac{|\textbf k_{q_i}|}{|\textbf k_{\lambda_i}|}\sqrt{\frac{|m w_0|-2 m^2 w_0 w_2\cos\phi_r}{3+ 4m^2 w_2^2}}\Bigg]},
    \end{equation}
    and is estimated to be approximately $21.59^\circ$ based on the fitting parameters obtained in the previous section.
    At this magic angle, the bandwidth is given by
    \begin{align}
        \Delta E_2=\Delta E_3=&\frac{1}{6w_2^2+8m^2 w_2^4}\bigg[w_0(1-4m^2 w_2^2)(4-2m w_2\cos\phi_r+4m^2 w_2^2)\nonumber\\
        &+\sqrt{16(1-m w_2\cos\phi_r+2m^2 w_2^2)-4m^2 w_2^2(1-4m^2 w_2^2)\sin^2{\phi_r}})\bigg].
    \end{align}

    \begin{figure}[H]
        \centering
  	\includegraphics[width=0.6\linewidth]{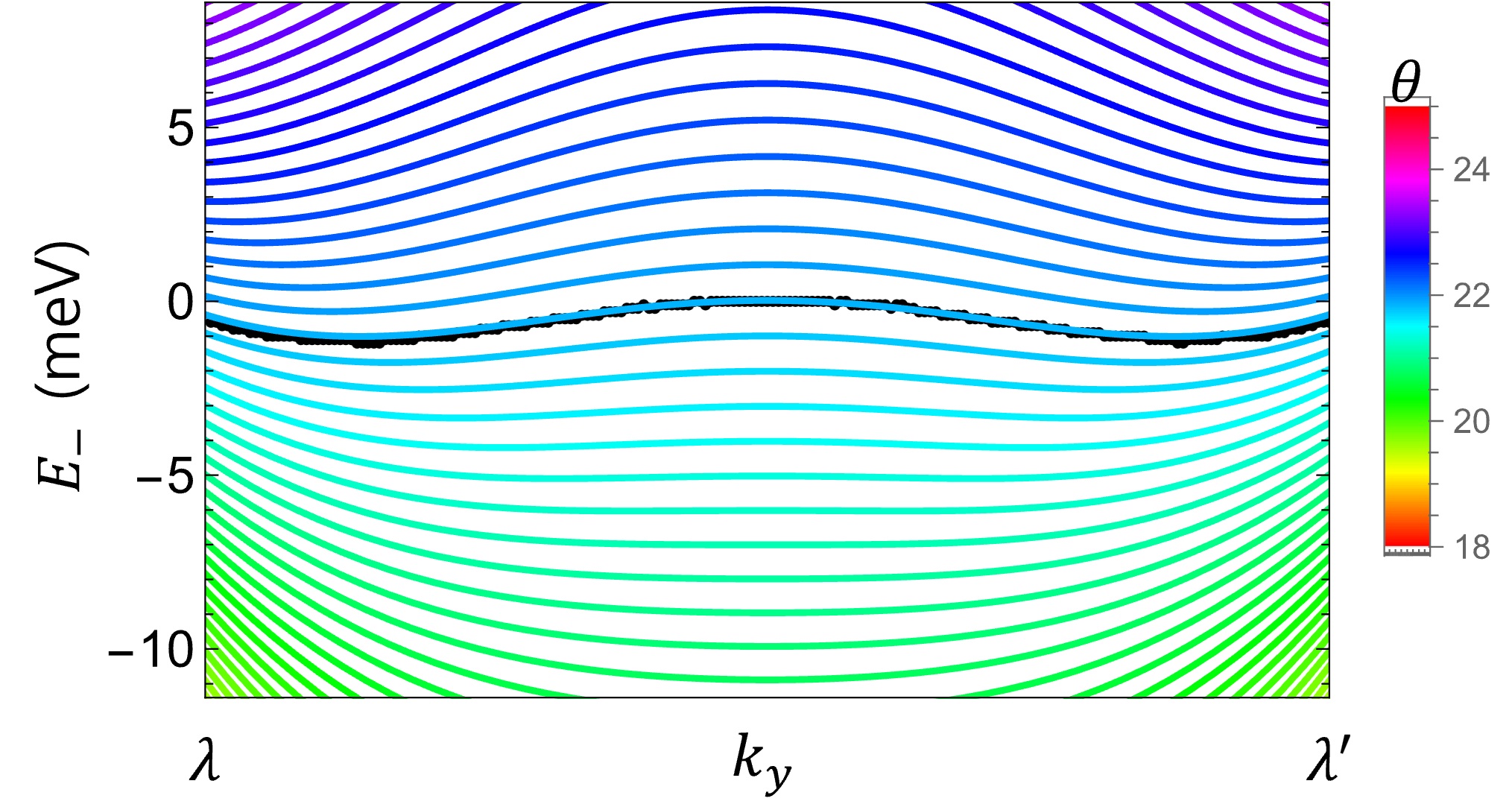}
  	\caption{
        {\bf {Numerically calculated band dispersion as a function of twist angle $\theta$.}}
  		The black dots show the dispersion obtained from the first principles calculations.
  	}\label{Fig4}
    \end{figure}

    While the twist-angle dependence was encoded only in $k_{0y}$ in our analysis above, other quantities—such as the effective mass along the $\lambda-\lambda'$ direction and the band minima shift along the $\gamma-\kappa$ ($\gamma-\kappa'$) direction—are in fact also $\theta$-dependent. This is because, as shown in Eq.~\eqref{H12}, $\theta$ enters the diagonal elements of the Hamiltonian explicitly through $R(\pm \theta/2)$, $\textbf a_{q_1}$, and $\textbf a_{q_2}$. Keeping only $w_0$, $w_2$, and $\phi_r$ fixed, we compute the dispersions for twist angles near $21.8^\circ$ [Fig.~\ref{Fig4}] and compare the numerically obtained $\Delta E_1$ with the analytic result. As shown in Fig 5c of the main text, the discrepancy between the analytic approximation and the numerical calculation increases as $\theta$ deviates from the reference angle of $21.8^\circ$. Nevertheless, $\Delta E_1$ remains a reliable indicator of the bandwidth, and the magic angle is still found to lie very close to $21.8^\circ$. From the numerical analysis that considers the twist-angle dependence of the effective mass and band minima shift, the magic twist angle corresponding to the minimum bandwidth is found to be $\theta_{c,\text{num}}=21.7^\circ$.

    The magic angle discussed above corresponds to the twist angle that minimizes the bandwidth. We note, however, that one can define another type of magic angle from the perspective of the effective mass. Specifically, we find that the second derivative of $E_-(k_y)$ in Eq.~\eqref{epm} vanishes at $k_y=0$ when $\theta=\theta_{c2}$, where
    \begin{equation}\label{secang}
        \theta_{c2}=2\arcsin{\Bigg[\frac{3}{2\pi}\frac{|\textbf k_{q_i}|}{|\textbf k_{\lambda_i}|}\sqrt{|mw_0|-2m^2 w_0 w_2\cos\phi_r}\Bigg]}\approx20.39^\circ.
    \end{equation}
    This condition defines a secondary magic angle, at which the leading term in the expansion of $E_-(k_y)$ near $k_y=0$ becomes quartic. Under this circumstance, the density of states exhibits an exact power-law divergence. For details, see \textit{Power-law-divergent density of states from the taco band} section.

    \subsection{Band structure over the whole moir\'e Brillouin zone}
    In this section, we discuss the band structure and density of states (DOS) obtained by extending our continuum model beyond the line connecting $\lambda$ and $\lambda'$. As in Fig.~\ref{Fig5}, we assume that the interlayer tunneling parameters remain approximately constant as the twist angle varies.

    \textit{General momentum dependence of the interlayer tunneling}---
    To construct a more general model that covers the whole mBZ, we need to find the form of the interlayer tunneling $t(k_x,k_y)$ at a general $\textbf k$. For this we approximate that $t(k_x,k_y)$ has a polynomial form:
    \begin{equation}
        t(k_x,k_y)=\sum_{a,b=0}^{2}W_{ab}(k_x-k_{0x})^ak_y^b,
    \end{equation}
    where $k_{0x}=[\textbf k_{\lambda_1}+\textbf k_{\lambda_2}]_x/2$
    Here, as we assumed quadratic dispersion near the valleys, we cut the series of $t$ up to the second order in $k_x$ and $k_y$.
    Due to the three $C_{2y}T$-type symmetries, any $W_{ab}$ with odd $b$ must be zero:
    \begin{align}
        W_{ab}=0\text{ for $b=$ odd}.
    \end{align}
    From the numerical data fitting along the $\lambda-\lambda'$ line, we have
    \begin{align}
        W_{00}&=w_0,\\
        W_{02}&=w_2.
    \end{align}
    From the fitting along the $\gamma-\mu$ line where $k_y=0$, we find that
    \begin{align}
        W_{10}&\approx0.0235 a_0\text{ eV},\\
        W_{20}&\approx0.0347 a_0^2\text{ eV}.
    \end{align}
    The remaining $W_{ab}$ could in principle be obtained from the data fitting along the $\gamma-\lambda$ line. However, the lowest-order term with both $a\neq0$ and $b\neq0$ is already cubic ($a=1, b=2$) in $k_x$ and $k_y$, is therefore  excluded from the model.

    \textit{The 12-band continuum model Hamiltonian}---
    Our model includes six $\lambda_i$ points. For each $\lambda_i$ at $\textbf k_{\lambda_i}\approx0.700~\textbf k_{q_i}$, there are two valleys, one with spin-up and the other with spin-down. The Hamiltonian reads
    \begin{equation}\label{Hud}
        H(\textbf k)=
        \begin{pmatrix}
            H_\uparrow(\textbf k) & \\
             & H_\downarrow(\textbf k)
        \end{pmatrix},
    \end{equation}
    where
    \begin{equation}\label{Hup}
        H_\uparrow(\textbf k)=
        \begin{pmatrix}
            H_{\uparrow1}(\textbf k) & t_{12}(\textbf k) & 0 & 0 & 0 & 0 \\
            t_{12}^*(\textbf k) & H_{\uparrow2}(\textbf k) & 0 & 0 & 0 & 0 \\
            0 & 0 & H_{\uparrow3}(\textbf k) & t_{34}(\textbf k) & 0 & 0 \\
            0 & 0 & t_{34}^*(\textbf k) & H_{\uparrow4}(\textbf k) & 0 & 0 \\
            0 & 0 & 0 & 0 & H_{\uparrow5}(\textbf k) & t_{56}(\textbf k) \\
            0 & 0 & 0 & 0 & t_{56}^*(\textbf k) & H_{\uparrow6}(\textbf k)
        \end{pmatrix}
    \end{equation}
    and
    \begin{equation}\label{Hdown}
        H_\downarrow(\textbf k)=
        \begin{pmatrix}
            H_{\downarrow1}(\textbf k) & 0 & 0 & 0 & 0 & t_{61}(\textbf k) \\
            0 & H_{\downarrow2}(\textbf k) & t_{23}(\textbf k) & 0 & 0 & 0 \\
            0 & t_{23}^*(\textbf k) & H_{\downarrow3}(\textbf k) & 0 & 0 & 0 \\
            0 & 0 & 0 & H_{\downarrow4}(\textbf k) & t_{45}(\textbf k) & 0 \\
            0 & 0 & 0 & t_{45}^*(\textbf k) & H_{\downarrow5}\textbf k) & 0 \\
            t_{61}^*(\textbf k) & 0 & 0 & 0 & 0 & H_{\downarrow6}(\textbf k).
        \end{pmatrix}
    \end{equation}
    Here, $H_i(\textbf k)$ includes the kinetic energy and moir\'e potential of electron near the $\lambda_i$ valley, and $t_{ij}(\textbf k)$ is the coupling between the $\lambda_i$ and $\lambda_j$ valley due to the interlayer tunneling.
    The diagonal elements in Eq.~\ref{Hup} and Eq.~\ref{Hdown} are given by
    \begin{align}
        H_{\uparrow i}{(\textbf k)}&=\sum_{\alpha=x,y}\frac{[R(-\frac{2\lfloor(i-1)/2\rfloor}{3}\pi+(-1)^i\frac{\theta}{2})(\textbf k-\textbf k_{q_i})]_\alpha^2}{2m_\alpha}+A\cos{(\textbf a_{q_i}\cdot\textbf k+\varphi)},
    \end{align}
    and
    \begin{align}
        H_{\downarrow i}{(\textbf k)}&=\sum_{\alpha=x,y}\frac{[R(\frac{1-2\lfloor i/2\rfloor}{3}\pi-(-1)^i\frac{\theta}{2})(\textbf k-\textbf k_{q_i})]_\alpha^2}{2m_\alpha}+A\cos{(\textbf a_{q_i}\cdot\textbf k+\varphi)},
    \end{align}
    where $\textbf a_i=[R(\pi/3)]^{(i-1)}(a_0/\sin{(\theta/2)})(-\sqrt{3}/4,1/4)$.
    The off-diagonal elements in Eq.~\ref{Hup} and Eq.~\ref{Hdown} are given by
    \begin{align}
        t_{ij}{(\textbf k)}&=w_0+w_2 e^{i\phi_r}q_{ij,2}^2(\textbf k)+W_{10}(q_{ij,1}(\textbf k)-k_{0x})+W_{20}(q_{ij,1}(\textbf k)-k_{0x})^2,\\
    \end{align}
    where
    \begin{align}
        q_{ij,1}{(\textbf k)}&=\textbf k\cdot(\hat z\times\hat n_{ij}),\\
        q_{ij,2}{(\textbf k)}&=\textbf k\cdot\hat n_{ij},\\
        \hat n_{ij}&\equiv\frac{\textbf k_{\lambda_j}-\textbf k_{\lambda_i}}{|\textbf k_{\lambda_j}-\textbf k_{\lambda_i}|}.
    \end{align}
    We note that $q_{12,1}(\textbf k)=k_x$ and $q_{12,2}(\textbf k)=k_y$.
    
    \subsection{Taco band structure}

    \begin{figure}[H]
        \centering
  	\includegraphics[width=0.8\linewidth]{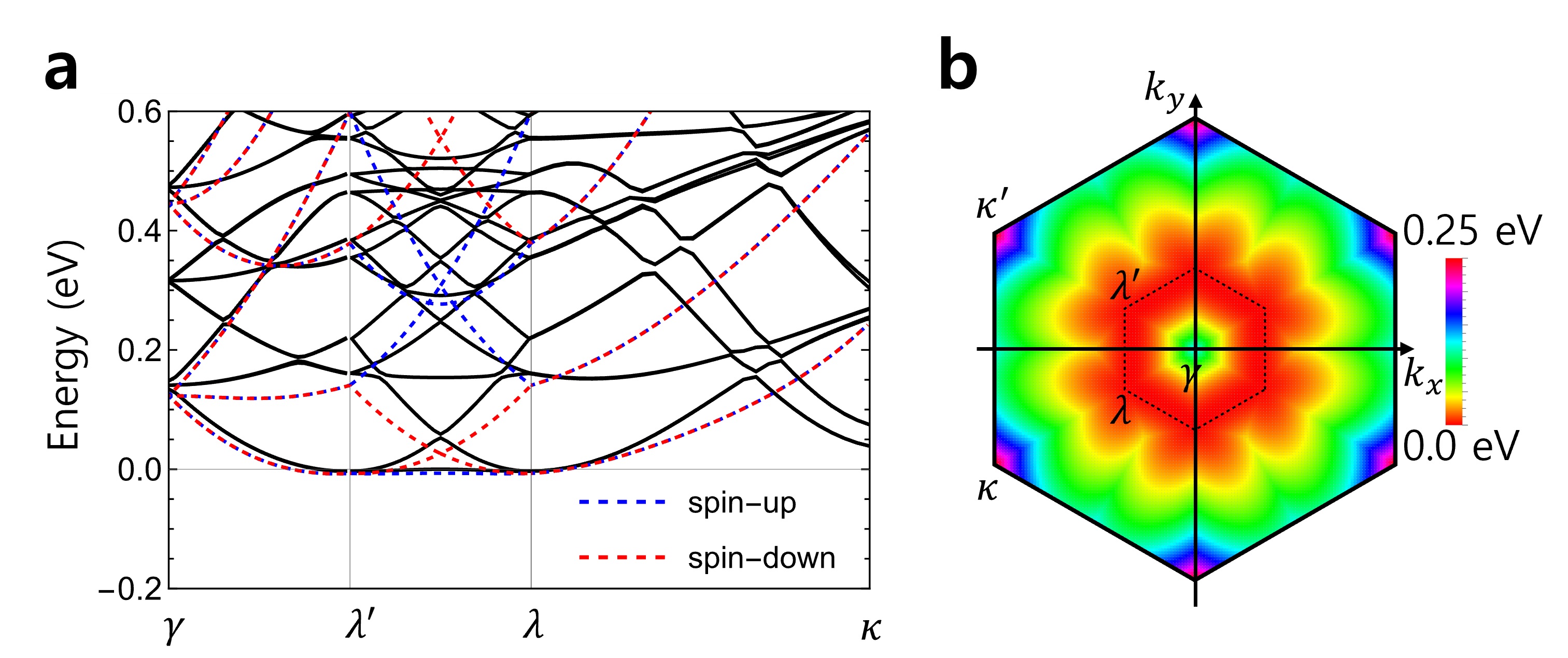}
  	\caption{
        {\bf {Band structure from the 12-band continuum model}.}
  		(\textbf{a}) The red and blue dashed lines show the band structures of spin-up and -down bands, respectively, obtained by diagonalizing the continuum model Hamiltonian. The black solid lines represent the results from first-principles calculations.
        (\textbf{b}) The energy spectrum of the lowest band (taco band).
  	}\label{band}
    \end{figure}
    
    Fig.~\ref{band}a shows the band structure obtained along the $\gamma-\lambda'-\lambda-\kappa$ path.
    Compared to the first-principles results, the model calculation reproduces well the dispersion of the lowest-energy bands of $H_\uparrow(\textbf k)$ and $H_\downarrow(\textbf k)$ around the $\lambda$ and $\lambda'$ points. However, it underestimates the bandwidths of the second and third bands, while overestimating the bandwidths of higher bands. This discrepancy arises because the higher-energy bands originate from different $\lambda_i$ valleys than those chosen as $\lambda$ and $\lambda'$, and the polynomial approximation of the Hamiltonian, truncated at second order in momentum, introduces errors at momenta far from the valleys.
    Fig.~\ref{band}b shows the energy spectrum of the taco band.
    Here, “taco” refers to the fact that along the edges of the hexagon formed by the $\lambda_i$ points, the band dispersion is nearly flat, while it is quadratic in the perpendicular direction, resembling the shape of a folded taco.
    
    \textit{Power-law-divergent density of states from the taco band}---
    The DOS shown in Fig. 5d of the main text is computed from
    \begin{align}
        D(E)&=\frac{1}{N}\sum_{i=1}^{12}\sum_{\textbf k\in\textrm{mBZ}} \delta(E-\varepsilon_i(\textbf k))\nonumber\\
        &\approx\frac{1}{\pi N}\sum_{i=1}^{12}\sum_{\textbf k\in\textrm{mBZ}} \frac{\eta}{(E-\varepsilon_i(\textbf k))^2+\eta^2},
    \end{align}
    where $\varepsilon_i(\textbf k)$ denote the $i$-th eigenvalue of $H(\textbf k)$ in Eq.~\ref{Hud} and $N$ is the system size.
    To numerically calculate DOS, we introduce a small broadening factor $\eta$.
    We find that the DOS of magic-angle twisted bilayer WSe$2$ exhibits a divergence near the energy of the taco band. This divergence originates from a van Hove singularity (vHS)~\cite{van1953occurrence}, corresponding to a saddle point in the band dispersion located at the midpoint of the $\lambda$–$\lambda'$ line with energy $E_-(0)$.
    Consequently, the DOS $D(E)$ diverges logarithmically, $D(E)\propto -\log_{10}(E-E_-(0))$ in the vicinity of $E_-(0)$.
    However, this strictly logarithmic behavior is visible only within an extremely narrow energy window set by the taco band’s bandwidth.
    At larger energy scales, the DOS instead appears to follow an approximate power-law divergence, $D(E)\propto (E-E_-(0))^{-a}$.

\begin{figure}[H]
        \centering
  	\includegraphics[width=1.\linewidth]{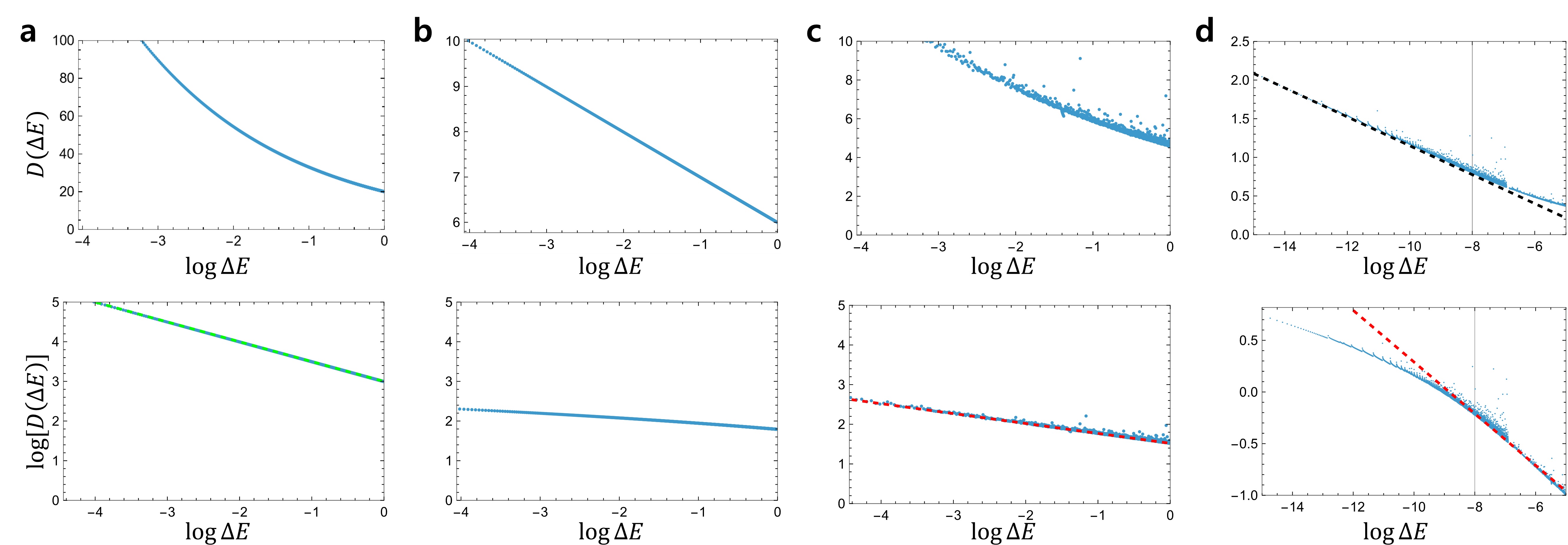}
  	\caption{
        {\bf {DOS from the model dispersion in Eq.~\eqref{smodiv}}}
  		Density of states calculated from Eq.\eqref{modos} with $p=4$ for (\textbf{a}) $a=1$, $b=c=0$, (\textbf{b}) $a=1$, $b=-1$, $c=0$, (\textbf{c}) $a=1$, $b=0$, $c=0.1$, and (\textbf{d}) $a=0.54777$, $b=-0.03699$, $c=0.61191$.
        The parameters used for (\textbf{d}) approximately describes the dispersion in the vicinity of the van Hove singularity on the $\lambda-\lambda'$ line.
        The upper (lower) panels show the $\log_{10}D(\Delta E)$ vs. $\log_{10}\Delta E$ ($D(\Delta E)$ vs. $\log_{10}\Delta E$) plots. In the logarithmically divergent case, the $D(\Delta E)$ vs. $\log_{10}\Delta E$ plot becomes linear, whereas in the power-law case, the $\log_{10}D(\Delta E)$ vs. $\log_{10}\Delta E$ plot becomes linear. The colored dashed lines in (\textbf{a}), (\textbf{c}), and (\textbf{d}) indicate reference power laws with exponents $-1/2$ (green) and $-1/4$ (red). The black dashed line shows that the divergent behavior of (\textbf{d}) is apparently logarithmic at low energy scales.
        The scattered points in (\textbf{c}) and (\textbf{d}) are due to numerical integration errors.
  	}\label{div}
    \end{figure}

    This power-law divergence originates from the quasi-one-dimensional (1D) nature of the taco band. In 1D, a vHS corresponds to an extremum in the band dispersion, and it leads to a power-law divergence of the DOS. Along the $\lambda$–$\lambda'$ lines, the taco band is nearly flat in one direction while it exhibits a minimum in the other direction, making it effectively a quasi-1D vHS.
    Notably, the exponent $a$ of DOS in Fig. 5e of the main text is smaller than the ideal 1D value of $1/2$. This is because the taco band remains quasi-1D flat only along short line segments, while becoming dispersive away from the $\lambda$–$\lambda'$ line.
    To capture this, we consider the model dispersion
    \begin{equation}\label{smodiv}
      \varepsilon(\textbf k)=\varepsilon_0+ak_x^2+bk_y^2+ck_y^p,
    \end{equation}
    with $p$ an even integer larger than 2. The DOS is then
    \begin{align}\label{modos}
        D(\Delta E)&=\int\frac{d^2k}{(2\pi)^2}\delta(E-\varepsilon(\textbf k))\nonumber\\
        &=\frac{1}{(2\pi)^2}\int_{-\infty}^{\infty}dk_y\frac{\Theta(\Delta E-bk_y^2-ck_y^p)}{\sqrt{a(\Delta E-bk_y^2-ck_y^p)}},
    \end{align}
    where $\Delta E=E-\varepsilon_0$ and $\Theta$ is the Heaviside step function.
    Different parameter choices yield distinct types of singularities. When $b=c=0$, the dispersion is perfectly 1D, and $D(\Delta E)\propto 1/\sqrt{\Delta E}$. For $b\gg c>0$, the DOS does not diverge. For $b<0$, $(k_x,k_y)=(0,0)$ becomes a saddle point, giving a logarithmic divergence, $D(\Delta E)\propto -\log_{10}\Delta E$. For $b=0$ and $c>0$, one finds $D(\Delta E)\propto (\Delta E)^{-(1/2-1/p)}$; in particular, for $p=4$, the DOS diverges with exponent $1/4$.
    Thus, depending on $b$, $c$, and $p$, the DOS at $\Delta E=0$ exhibits either no divergence, logarithmic divergence, or a power-law divergence with exponent less than $1/2$. For small negative $b$ and positive $c$, the divergence is intermediate between logarithmic and the $1/4$ power law. This explains why quasi-1D band structures, such as the taco band, can exhibit approximate power-law divergences with exponents significantly smaller than $1/2$. Fig.~\ref{div} shows $D(\Delta E)$ vs. $\log_{10}\Delta E$ and $\log_{10}D(\Delta E)$ vs. $\log_{10}\Delta E$ graphs for different $b$ and $c$ values with $a=1$ and $p=4$.
    For $a$, $b$, and $c$ extracted from the saddle point in the dispersion of the Hamiltonian in Eq.~\eqref{Hud} with $p=4$, the $\log_{10}D(\Delta E)$ vs. $\log_{10}\Delta E$ plot appears linear at larger energy scales, whereas the $D(\Delta E)$ vs. $\log_{10}\Delta E$ plot becomes linear at smaller energy scales. This crossover behavior is consistent with the DOS calculated from the model Hamiltonian in Eq.~\eqref{Hud}, and the estimated crossover energy scale, $\Delta E\approx 0.3$ meV, shows good agreement between the two results.

\begin{figure}[H]
        \centering
  	\includegraphics[width=1.\linewidth]{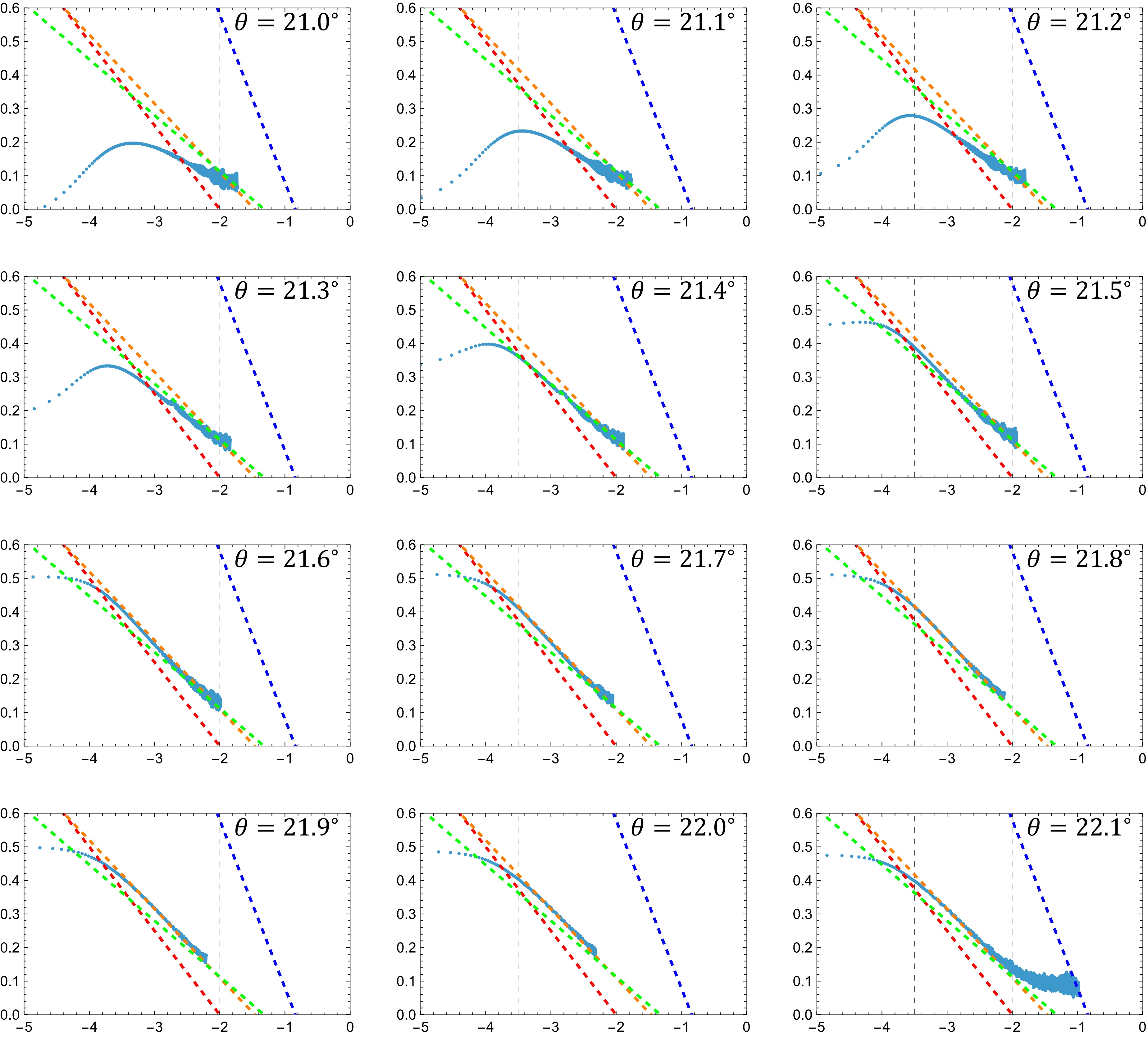}
  	\caption{
  		{\bf {$\log_{10}D(\Delta E)$ vs. $\log_{10}\Delta E$ at $\theta$ from 21.0$^\circ$ to 22.1$^\circ$ with $\eta=0.1$ meV.}}
        The blue, red, orange, and green dashed lines indicate the reference power laws with exponents $-1/2$, $-1/4$, $-0.205$, and $-1/6$. The vertical dashed lines denote the energy range within which the power-law divergence appears at 21.8$^\circ$.
  	}\label{sangles}
    \end{figure}

    Finally, we comment on the dependence of the DOS on the twist angle $\theta$, the broadening factor $\eta$, and the nature of the divergence at the secondary magic angle.
    Fig.~\ref{sangles} shows the log–log DOS at several twist angles near $21.8^\circ$, computed with $\eta=0.1$ meV. For $\theta\sim 21^\circ$, no clear divergence is observed, although a peak structure in the DOS is already visible. As $\theta$ increases, the peak sharpens, and once $\theta > 21.4^\circ$, the midpoint of the $\lambda$–$\lambda'$ line evolves into a saddle point. The DOS reaches its maximum near $\theta=21.7^\circ$–$21.8^\circ$, where the taco-band bandwidth is minimized. For larger $\theta$, however, the energy window supporting approximate power-law divergence shrinks.

    Numerically, the secondary magic angle—defined by the vanishing effective mass at the midpoint of the $\lambda$–$\lambda'$ line—is found to be $\theta_{c2}\approx 21.35^\circ$, slightly larger than the analytic prediction of Eq.~\eqref{secang}. This secondary magic angle acts as a critical point: for smaller $\theta$, a finite DOS peak exists without true divergence, while for larger $\theta$, a van Hove singularity emerges accompanied by an extremely large effective mass along one direction.

    From theory, the predicted divergence exponent at the secondary magic angle is $-1/4$, since the dispersion follows a quartic form in momentum near the band minimum. In the calculations above, however, the observed divergence at the secondary magic angle is even weaker than at the primary magic angle, primarily due to the finite broadening factor. As illustrated in Fig.~\ref{sangles}, the DOS at $\theta=21.35^\circ$ with much smaller broadening factor $\eta=0.002$ meV indeed exhibits a power-law divergence with exponent close to $-1/4$ at very low energies, but this behavior is rapidly cut off once $\Delta E$ approaches the broadening scale.
    We further note that the energy range of approximate power-law scaling is much wider at $\theta=21.8^\circ$ than at $\theta=21.35^\circ$. This happens because the quartic ($k_y^4$) approximation to the dispersion is valid only within a limited momentum window. At larger $k_y$, the dispersion crosses over to quadratic behavior. Near the primary magic angle, the flattening extends over a broader momentum range, thereby sustaining the $k_y^4$ approximation across a wider energy window. In contrast, near the secondary magic angle, the dispersion may locally resemble a $k_y^4$ form almost perfectly, but only over a very narrow $k_y$-range. As a result, the effective power-law scaling breaks down much more quickly, limiting the observable regime for divergence.
    
    \begin{figure}[H]
        \centering
  	\includegraphics[width=0.4\linewidth]{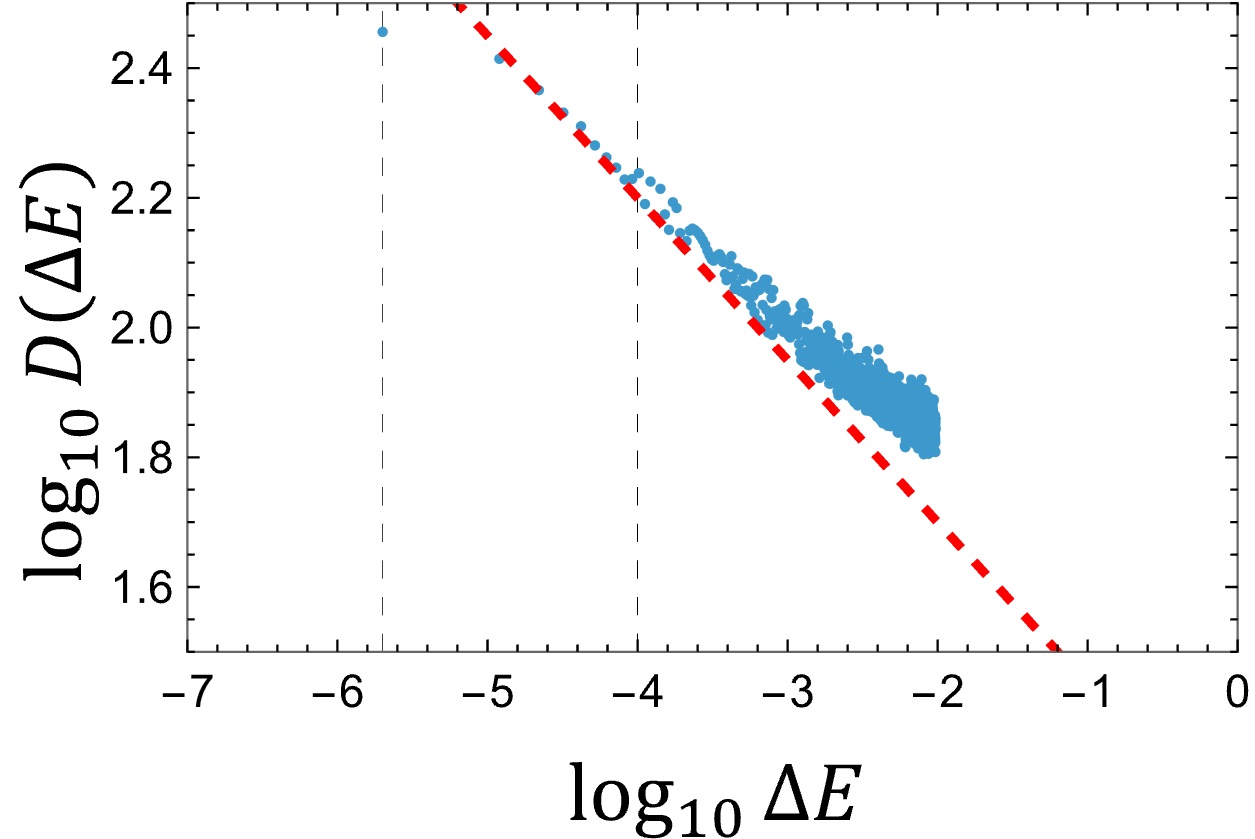}
  	\caption{
        {\bf {$\log_{10}D(\Delta E)$ vs. $\log_{10}\Delta E$ at $\theta=21.35^\circ$ with $\eta=0.002$ meV.}}
  		The red dashed line indicates the reference power law with an exponent $-1/4$. The vertical dashed line denotes the broadening factor size.
  	}\label{secmag}
    \end{figure}

\begin{figure}[H]
        \centering
  	\includegraphics[width=0.6\linewidth]{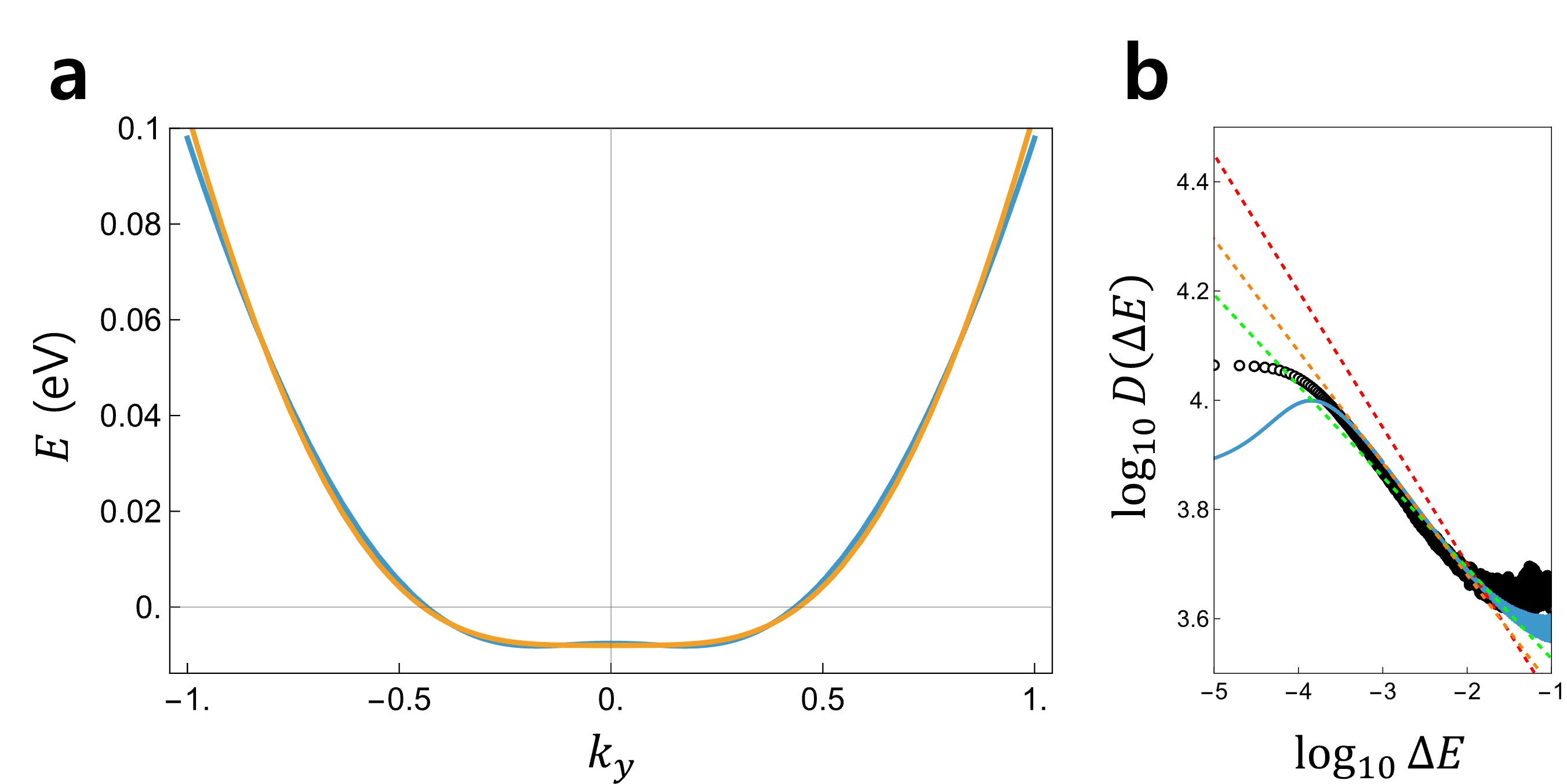}
  	\caption{
        {\bf {Calculations for the $ak_x^2+bk_y^2\tanh^2(\beta k_y)$ dispersion.}}
  		(\textbf{a}) Band dispersion along the extended $\lambda$–$\lambda'$ direction. The blue line shows the result from the full model Hamiltonian, while the orange line shows the analytic dispersion $a k_x^2 + b k_y^2 \tanh^2(\beta k_y)$ with $a \approx 0.545777$, $b \approx 1.5$, and $\beta \approx 1.3$.
        (\textbf{b}) $\log_{10} D(\Delta E)$ vs. $\log_{10} \Delta E$ obtained from the analytic dispersion with the same parameters. The red, orange, and green dashed lines indicate reference power laws with exponents $-1/4$, $-0.205$, and $-1/6$, respectively. Black circles represent the corresponding results from the full model Hamiltonian.
  	}\label{tanhdos}
    \end{figure}

    To better capture the behavior of the taco band over a larger energy range, one can consider a dispersion of the form
    \begin{equation}\label{modiv2}
        \varepsilon(\mathbf{k})=ak_x^2+bk_y^2\tanh^2(\beta k_y).
    \end{equation}
    For small $k_y$, $\tanh(\beta k_y) \approx \beta k_y$, yielding a quartic dependence on $k_y$ ($\varepsilon(\mathbf{k}) \approx a k_x^2 + c k_y^4$ with $c = b \beta^2$). For large $k_y$, $\tanh(\beta k_y) \approx 1$, recovering a quadratic dispersion along $k_y$ ($\varepsilon(\mathbf{k}) \approx a k_x^2 + b k_y^2$). This model therefore captures the crossover from quartic to quadratic behavior and mimics the taco-band dispersion of our system over a larger energy window.
    Fig.~\ref{tanhdos}a shows the band dispersions from $a k_x^2 + b k_y^2 \tanh^2(\beta k_y)$ with parameters $a \approx 0.545777$, $b \approx 1.5$, and $\beta \approx 1.3$.
    Fig.~\ref{tanhdos}b presents the corresponding DOS from the analytic dispersion, which reproduces both the divergent exponent and the energy range of divergence seen in the full model.
    Since the DOS from the full model Hamiltonian is normalized differently (with units of eV$^{-1}$/unit cell/atom), we applied an overall multiplicative constant to align the two curves for visual comparison.

\section*{References}
\putbib[references]
\end{bibunit}


\begin{thebibliography}{10}

\bibitem{alexeev2019resonantly}
Evgeny~M. Alexeev, David~A. Ruiz-Tijerina, Mark Danovich, Matthew~J. Hamer, Daniel~J. Terry, Pramoda~K. Nayak, Seongjoon Ahn, Sangyeon Pak, Juwon Lee, Jung~Inn Sohn, Maciej~R. Molas, Maciej Koperski, Kenji Watanabe, Takashi Taniguchi, Kostya~S. Novoselov, Roman~V. Gorbachev, Hyeon~Suk Shin, Vladimir~I. Fal’ko, and Alexander~I. Tartakovskii.
\newblock {Resonantly hybridized excitons in moiré superlattices in van der Waals heterostructures}.
\newblock {\em Nature}, 567(7746):81--86, 3 2019.

\bibitem{angeli2021gamma}
Mattia Angeli and Allan~H. MacDonald.
\newblock {$\Gamma$ valley transition metal dichalcogenide moiré bands}.
\newblock {\em Proceedings of the National Academy of Sciences}, 118(10), 3 2021.

\bibitem{balents2020superconductivity}
Leon Balents, Cory~R. Dean, Dmitri~K. Efetov, and Andrea~F. Young.
\newblock {Superconductivity and strong correlations in moiré flat bands}.
\newblock {\em Nature Physics}, 16(7):725--733, 5 2020.

\bibitem{berg2012electronic}
Erez Berg, Mark~S Rudner, and Steven~A Kivelson.
\newblock Electronic liquid crystalline phases in a spin-orbit coupled two-dimensional electron gas.
\newblock {\em Physical Review B—Condensed Matter and Materials Physics}, 85(3):035116, 2012.

\bibitem{bistritzer2010transport}
R~Bistritzer and AH~MacDonald.
\newblock Transport between twisted graphene layers.
\newblock {\em Physical Review B}, 81(24):245412, 2010.

\bibitem{bistritzer2011moire}
Rafi Bistritzer and Allan~H. MacDonald.
\newblock {Moiré bands in twisted double-layer graphene}.
\newblock {\em Proceedings of the National Academy of Sciences}, 108(30):12233--12237, 7 2011.

\bibitem{bradley2015probing}
Aaron~J. Bradley, Miguel~M. Ugeda, Felipe~H. Da~Jornada, Diana~Y. Qiu, Wei Ruan, Yi~Zhang, Sebastian Wickenburg, Alexander Riss, Jiong Lu, Sung-Kwan Mo, Zahid Hussain, Zhi-Xun Shen, Steven~G. Louie, and Michael~F. Crommie.
\newblock {Probing the role of interlayer coupling and coulomb interactions on electronic structure in Few-Layer MOSE2 nanostructures}.
\newblock {\em Nano Letters}, 15(4):2594--2599, 3 2015.

\bibitem{cai2023signatures}
Jiaqi Cai, Eric Anderson, Chong Wang, Xiaowei Zhang, Xiaoyu Liu, William Holtzmann, Yinong Zhang, Fengren Fan, Takashi Taniguchi, Kenji Watanabe, Ying Ran, Ting Cao, Liang Fu, Di~Xiao, Wang Yao, and Xiaodong Xu.
\newblock {Signatures of fractional quantum anomalous Hall states in twisted MoTe2}.
\newblock {\em Nature}, 622(7981):63--68, 6 2023.

\bibitem{cao2018correlated}
Yuan Cao, Valla Fatemi, Ahmet Demir, Shiang Fang, Spencer~L. Tomarken, Jason~Y. Luo, Javier~D. Sanchez-Yamagishi, Kenji Watanabe, Takashi Taniguchi, Efthimios Kaxiras, Ray~C. Ashoori, and Pablo Jarillo-Herrero.
\newblock {Correlated insulator behaviour at half-filling in magic-angle graphene superlattices}.
\newblock {\em Nature}, 556(7699):80--84, 3 2018.

\bibitem{cao2018unconventional}
Yuan Cao, Valla Fatemi, Shiang Fang, Kenji Watanabe, Takashi Taniguchi, Efthimios Kaxiras, and Pablo Jarillo-Herrero.
\newblock {Unconventional superconductivity in magic-angle graphene superlattices}.
\newblock {\em Nature}, 556(7699):43--50, 3 2018.

\bibitem{carr2019exact}
Stephen Carr, Shiang Fang, Ziyan Zhu, and Efthimios Kaxiras.
\newblock {Exact continuum model for low-energy electronic states of twisted bilayer graphene}.
\newblock {\em Physical Review Research}, 1(1), 8 2019.

\bibitem{carr2017twistronics}
Stephen Carr, Daniel Massatt, Shiang Fang, Paul Cazeaux, Mitchell Luskin, and Efthimios Kaxiras.
\newblock {Twistronics: Manipulating the electronic properties of two-dimensional layered structures through their twist angle}.
\newblock {\em Physical review. B./Physical review. B}, 95(7), 2 2017.

\bibitem{carr2018relaxation}
Stephen Carr, Daniel Massatt, Steven~B. Torrisi, Paul Cazeaux, Mitchell Luskin, and Efthimios Kaxiras.
\newblock {Relaxation and domain formation in incommensurate two-dimensional heterostructures}.
\newblock {\em Physical review. B./Physical review. B}, 98(22), 12 2018.

\bibitem{castro2008low}
Eduardo~V Castro, NMR Peres, T~Stauber, and NAP Silva.
\newblock Low-density ferromagnetism in biased bilayer graphene.
\newblock {\em Physical Review Letters}, 100(18):186803, 2008.

\bibitem{deng2025epitaxially}
Bingchen Deng, Heonsu Ahn, Jue Wang, Gunho Moon, Cheolhee Han, Ninad Dongre, Chao Lei, Giovanni Scuri, Jiho Sung, Elise Brutschea, et~al.
\newblock Epitaxially defined luttinger liquids on mos 2 bicrystals.
\newblock {\em Physical Review Letters}, 134(4):046301, 2025.

\bibitem{devakul2021magic}
Trithep Devakul, Valentin Crépel, Yang Zhang, and Liang Fu.
\newblock {Magic in twisted transition metal dichalcogenide bilayers}.
\newblock {\em Nature Communications}, 12(1), 11 2021.

\bibitem{drummond2009phase}
ND~Drummond and RJ~Needs.
\newblock Phase diagram of the low-density two-dimensional homogeneous electron gas.
\newblock {\em Physical review letters}, 102(12):126402, 2009.

\bibitem{du2023crossed}
X~Du, L~Kang, YY~Lv, JS~Zhou, X~Gu, RZ~Xu, QQ~Zhang, ZX~Yin, WX~Zhao, YD~Li, et~al.
\newblock Crossed luttinger liquid hidden in a quasi-two-dimensional material.
\newblock {\em Nature Physics}, 19(1):40--45, 2023.

\bibitem{enaldiev2020stacking}
VV~Enaldiev, Viktor Z{\'o}lyomi, CELAL Yelgel, SJ~Magorrian, and VI~Fal’Ko.
\newblock Stacking domains and dislocation networks in marginally twisted bilayers of transition metal dichalcogenides.
\newblock {\em Physical review letters}, 124(20):206101, 2020.

\bibitem{ghazaryan2021unconventional}
Areg Ghazaryan, Tobias Holder, Maksym Serbyn, and Erez Berg.
\newblock Unconventional superconductivity in systems with annular fermi surfaces: Application to rhombohedral trilayer graphene.
\newblock {\em Physical review letters}, 127(24):247001, 2021.

\bibitem{giamarchi2004theoretical}
T~Giamarchi.
\newblock Theoretical framework for quasi-one dimensional systems.
\newblock {\em Chemical reviews}, 104(11):5037--5056, 2004.

\bibitem{hou2022quantification}
Yaoping Hou, Guorui Wang, Chengfu Ma, Zhihua Feng, Yuhang Chen, and Tobin Filleter.
\newblock Quantification of the dielectric constant of mos2 and wse2 nanosheets by electrostatic force microscopy.
\newblock {\em Materials Characterization}, 193:112313, 2022.

\bibitem{hu2023light}
Chen Hu, Mit~H. Naik, Yang-Hao Chan, Jiawei Ruan, and Steven~G. Louie.
\newblock {Light-induced shift current vortex crystals in moiré heterobilayers}.
\newblock {\em Proceedings of the National Academy of Sciences}, 120(51), 12 2023.

\bibitem{hutchinson2016rashba}
Joel Hutchinson and Joseph Maciejko.
\newblock Rashba scattering in the low-energy limit.
\newblock {\em Physical Review B}, 93(24):245309, 2016.

\bibitem{hybertsen1986electron}
Mark~S. Hybertsen and Steven~G. Louie.
\newblock {Electron correlation in semiconductors and insulators: Band gaps and quasiparticle energies}.
\newblock {\em Physical review. B, Condensed matter}, 34(8):5390--5413, 10 1986.

\bibitem{jia2023tetragonal}
Ningning Jia, Yongting Shi, Zhiheng Lv, Junting Qin, Jiangtao Cai, Xue Jiang, Jijun Zhao, and Zhifeng Liu.
\newblock Tetragonal mexican-hat dispersion and switchable half-metal state with multiple anisotropic weyl fermions in penta-graphene.
\newblock {\em New Journal of Physics}, 25(3):033033, 2023.

\bibitem{jiang2020van}
Wenchao Jiang, Bowen Li, Xiaomeng Wang, Guanyu Chen, Tong Chen, Ying Xiang, Wei Xie, Yaomin Dai, Xiyu Zhu, Huan Yang, et~al.
\newblock Van hove singularity arising from mexican-hat-shaped inverted bands in the topological insulator sn-doped bi 1.1 sb 0.9 te 2 s.
\newblock {\em Physical Review B}, 101(12):121115, 2020.

\bibitem{kang2021band}
L~Kang, X~Du, JS~Zhou, X~Gu, YJ~Chen, RZ~Xu, QQ~Zhang, SC~Sun, ZX~Yin, YW~Li, et~al.
\newblock Band-selective holstein polaron in luttinger liquid material a 0.3 moo3 (a= k, rb).
\newblock {\em Nature Communications}, 12(1):6183, 2021.

\bibitem{krempasky2024altermagnetic}
J~Krempask{\`y}, L~{\v{S}}mejkal, SW~D’souza, M~Hajlaoui, G~Springholz, K~Uhl{\'\i}{\v{r}}ov{\'a}, F~Alarab, PC~Constantinou, V~Strocov, D~Usanov, et~al.
\newblock Altermagnetic lifting of kramers spin degeneracy.
\newblock {\em Nature}, 626(7999):517--522, 2024.

\bibitem{lee2024broken}
Suyoung Lee, Sangjae Lee, Saegyeol Jung, Jiwon Jung, Donghan Kim, Yeonjae Lee, Byeongjun Seok, Jaeyoung Kim, Byeong~Gyu Park, Libor {\v{S}}mejkal, et~al.
\newblock Broken kramers degeneracy in altermagnetic mnte.
\newblock {\em Physical review letters}, 132(3):036702, 2024.

\bibitem{li2021imagingmoire}
Hongyuan Li, Shaowei Li, Mit~H. Naik, Jingxu Xie, Xinyu Li, Jiayin Wang, Emma Regan, Danqing Wang, Wenyu Zhao, Sihan Zhao, Salman Kahn, Kentaro Yumigeta, Mark Blei, Takashi Taniguchi, Kenji Watanabe, Sefaattin Tongay, Alex Zettl, Steven~G. Louie, Feng Wang, and Michael~F. Crommie.
\newblock {Imaging moiré flat bands in three-dimensional reconstructed WSe2/WS2 superlattices}.
\newblock {\em Nature Materials}, 20(7):945--950, 2 2021.

\bibitem{li2021imaging}
Hongyuan Li, Shaowei Li, Emma~C. Regan, Danqing Wang, Wenyu Zhao, Salman Kahn, Kentaro Yumigeta, Mark Blei, Takashi Taniguchi, Kenji Watanabe, Sefaattin Tongay, Alex Zettl, Michael~F. Crommie, and Feng Wang.
\newblock {Imaging two-dimensional generalized Wigner crystals}.
\newblock {\em Nature}, 597(7878):650--654, 9 2021.

\bibitem{li2024imaging}
Hongyuan Li, Ziyu Xiang, Mit~H. Naik, Woochang Kim, Zhenglu Li, Renee Sailus, Rounak Banerjee, Takashi Taniguchi, Kenji Watanabe, Sefaattin Tongay, Alex Zettl, Felipe~H. Da~Jornada, Steven~G. Louie, Michael~F. Crommie, and Feng Wang.
\newblock {Imaging moiré excited states with photocurrent tunnelling microscopy}.
\newblock {\em Nature Materials}, 23(5):633--638, 1 2024.

\bibitem{li2024wigner}
Hongyuan Li, Ziyu Xiang, Aidan~P. Reddy, Trithep Devakul, Renee Sailus, Rounak Banerjee, Takashi Taniguchi, Kenji Watanabe, Sefaattin Tongay, Alex Zettl, Liang Fu, Michael~F. Crommie, and Feng Wang.
\newblock {Wigner molecular crystals from multielectron moiré artificial atoms}.
\newblock {\em Science}, 385(6704):86--91, 7 2024.

\bibitem{li2024imagingtunable}
Hongyuan Li, Ziyu Xiang, Tianle Wang, Mit~H. Naik, Woochang Kim, Jiahui Nie, Shiyu Li, Zhehao Ge, Zehao He, Yunbo Ou, Rounak Banerjee, Takashi Taniguchi, Kenji Watanabe, Sefaattin Tongay, Alex Zettl, Steven~G. Louie, Michael~P. Zaletel, Michael~F. Crommie, and Feng Wang.
\newblock {Imaging tunable Luttinger liquid systems in van der Waals heterostructures}.
\newblock {\em Nature}, 631(8022):765--770, 7 2024.

\bibitem{li2021quantum}
Tingxin Li, Shengwei Jiang, Bowen Shen, Yang Zhang, Lizhong Li, Zui Tao, Trithep Devakul, Kenji Watanabe, Takashi Taniguchi, Liang Fu, Jie Shan, and Kin~Fai Mak.
\newblock {Quantum anomalous Hall effect from intertwined moiré bands}.
\newblock {\em Nature}, 600(7890):641--646, 12 2021.

\bibitem{li2025robust}
Yanxing Li, Chuqiao Shi, Fan Zhang, Xiaohui Liu, Yuan Xue, Viet-Anh Ha, Qiang Gao, Chengye Dong, Yu-Chuan Lin, Luke~N Holtzman, et~al.
\newblock Robust supermoir{\'e} pattern in large-angle single-twist bilayers.
\newblock {\em Nature Physics}, pages 1--8, 2025.

\bibitem{li2024tuning}
Yanxing Li, Fan Zhang, Viet-Anh Ha, Yu-Chuan Lin, Chengye Dong, Qiang Gao, Zhida Liu, Xiaohui Liu, Sae~Hee Ryu, Hyunsue Kim, Chris Jozwiak, Aaron Bostwick, Kenji Watanabe, Takashi Taniguchi, Bishoy Kousa, Xiaoqin Li, Eli Rotenberg, Eslam Khalaf, Joshua~A. Robinson, Feliciano Giustino, and Chih-Kang Shih.
\newblock {Tuning commensurability in twisted van der Waals bilayers}.
\newblock {\em Nature}, 625(7995):494--499, 1 2024.

\bibitem{liu2021excitonic}
Erfu Liu, Takashi Taniguchi, Kenji Watanabe, Nathaniel~M. Gabor, Yong-Tao Cui, and Chun~Hung Lui.
\newblock {Excitonic and Valley-Polarization signatures of fractional correlated electronic phases in a WSE2/WS2 Moiré superlattice}.
\newblock {\em Physical Review Letters}, 127(3), 7 2021.

\bibitem{lu2024fractional}
Zhengguang Lu, Tonghang Han, Yuxuan Yao, Aidan~P. Reddy, Jixiang Yang, Junseok Seo, Kenji Watanabe, Takashi Taniguchi, Liang Fu, and Long Ju.
\newblock {Fractional quantum anomalous Hall effect in multilayer graphene}.
\newblock {\em Nature}, 626(8000):759--764, 2 2024.

\bibitem{luttinger1963exactly}
JM~Luttinger.
\newblock An exactly soluble model of a many-fermion system.
\newblock {\em Journal of mathematical physics}, 4(9):1154--1162, 1963.

\bibitem{maity2020phonons}
Indrajit Maity, Mit~H. Naik, Prabal~K. Maiti, H.~R. Krishnamurthy, and Manish Jain.
\newblock {Phonons in twisted transition-metal dichalcogenide bilayers: Ultrasoft phasons and a transition from a superlubric to a pinned phase}.
\newblock {\em Physical Review Research}, 2(1), 3 2020.

\bibitem{monarkha2012two}
Yu~P Monarkha and VE~Syvokon.
\newblock A two-dimensional wigner crystal.
\newblock {\em Low Temperature Physics}, 38(12):1067--1095, 2012.

\bibitem{naik2017origin}
Mit~H. Naik and Manish Jain.
\newblock {Origin of layer dependence in band structures of two-dimensional materials}.
\newblock {\em Physical review. B./Physical review. B}, 95(16), 4 2017.

\bibitem{naik2018ultraflatbands}
Mit~H. Naik and Manish Jain.
\newblock {Ultraflatbands and shear solitons in moiré patterns of twisted bilayer transition metal dichalcogenides}.
\newblock {\em Physical Review Letters}, 121(26), 12 2018.

\bibitem{naik2020origin}
Mit~H. Naik, Sudipta Kundu, Indrajit Maity, and Manish Jain.
\newblock {Origin and evolution of ultraflat bands in twisted bilayer transition metal dichalcogenides: Realization of triangular quantum dots}.
\newblock {\em Physical review. B./Physical review. B}, 102(7), 8 2020.

\bibitem{naik2022intralayer}
Mit~H. Naik, Emma~C. Regan, Zuocheng Zhang, Yang-Hao Chan, Zhenglu Li, Danqing Wang, Yoseob Yoon, Chin~Shen Ong, Wenyu Zhao, Sihan Zhao, M.~Iqbal~Bakti Utama, Beini Gao, Xin Wei, Mohammed Sayyad, Kentaro Yumigeta, Kenji Watanabe, Takashi Taniguchi, Sefaattin Tongay, Felipe~H. Da~Jornada, Feng Wang, and Steven~G. Louie.
\newblock {Intralayer charge-transfer moiré excitons in van der Waals superlattices}.
\newblock {\em Nature}, 609(7925):52--57, 8 2022.

\bibitem{park2023observation}
Heonjoon Park, Jiaqi Cai, Eric Anderson, Yinong Zhang, Jiayi Zhu, Xiaoyu Liu, Chong Wang, William Holtzmann, Chaowei Hu, Zhaoyu Liu, Takashi Taniguchi, Kenji Watanabe, Jiun-Haw Chu, Ting Cao, Liang Fu, Wang Yao, Cui-Zu Chang, David Cobden, Di~Xiao, and Xiaodong Xu.
\newblock {Observation of fractionally quantized anomalous Hall effect}.
\newblock {\em Nature}, 622(7981):74--79, 8 2023.

\bibitem{regan2020mott}
Emma~C. Regan, Danqing Wang, Chenhao Jin, M.~Iqbal~Bakti Utama, Beini Gao, Xin Wei, Sihan Zhao, Wenyu Zhao, Zuocheng Zhang, Kentaro Yumigeta, Mark Blei, Johan~D. Carlström, Kenji Watanabe, Takashi Taniguchi, Sefaattin Tongay, Michael Crommie, Alex Zettl, and Feng Wang.
\newblock {Mott and generalized Wigner crystal states in WSe2/WS2 moiré superlattices}.
\newblock {\em Nature}, 579(7799):359--363, 3 2020.

\bibitem{seixas2016multiferroic}
L~Seixas, AS~Rodin, A~Carvalho, and AH~Castro~Neto.
\newblock Multiferroic two-dimensional materials.
\newblock {\em Physical review letters}, 116(20):206803, 2016.

\bibitem{vsmejkal2022emerging}
Libor {\v{S}}mejkal, Jairo Sinova, and Tomas Jungwirth.
\newblock Emerging research landscape of altermagnetism.
\newblock {\em Physical Review X}, 12(4):040501, 2022.

\bibitem{splendiani2010emerging}
Andrea Splendiani, Liang Sun, Yuanbo Zhang, Tianshu Li, Jonghwan Kim, Chi-Yung Chim, Giulia Galli, and Feng Wang.
\newblock Emerging photoluminescence in monolayer mos2.
\newblock {\em Nano letters}, 10(4):1271--1275, 2010.

\bibitem{susarla2022hyperspectral}
Sandhya Susarla, Mit~H. Naik, Daria~D. Blach, Jonas Zipfel, Takashi Taniguchi, Kenji Watanabe, Libai Huang, Ramamoorthy Ramesh, Felipe~H. Da~Jornada, Steven~G. Louie, Peter Ercius, and Archana Raja.
\newblock {Hyperspectral imaging of exciton confinement within a moiré unit cell with a subnanometer electron probe}.
\newblock {\em Science}, 378(6625):1235--1239, 12 2022.

\bibitem{tanatar1989ground}
B~Tanatar and David~M Ceperley.
\newblock Ground state of the two-dimensional electron gas.
\newblock {\em Physical Review B}, 39(8):5005, 1989.

\bibitem{tillotson2024scanning}
Evan Tillotson, James~G McHugh, James Howarth, Teruo Hashimoto, Nicholas~J Clark, Astrid Weston, Vladimir Enaldiev, Sam Sullivan-Allsop, William Thornley, Wendong Wang, et~al.
\newblock Scanning electron microscopy imaging of twist domains in transition metal dichalcogenide heterostructures.
\newblock {\em ACS nano}, 18(50):34023--34033, 2024.

\bibitem{tomonaga1950remarks}
Sin-itiro Tomonaga.
\newblock Remarks on bloch's method of sound waves applied to many-fermion problems.
\newblock {\em Progress of Theoretical Physics}, 5(4):544--569, 1950.

\bibitem{tran2019evidence}
Kha Tran, Galan Moody, Fengcheng Wu, Xiaobo Lu, Junho Choi, Kyounghwan Kim, Amritesh Rai, Daniel~A. Sanchez, Jiamin Quan, Akshay Singh, Jacob Embley, André Zepeda, Marshall Campbell, Travis Autry, Takashi Taniguchi, Kenji Watanabe, Nanshu Lu, Sanjay~K. Banerjee, Kevin~L. Silverman, Suenne Kim, Emanuel Tutuc, Li~Yang, Allan~H. MacDonald, and Xiaoqin Li.
\newblock {Evidence for moiré excitons in van der Waals heterostructures}.
\newblock {\em Nature}, 567(7746):71--75, 2 2019.

\bibitem{turkel2022orderly}
Simon Turkel, Joshua Swann, Ziyan Zhu, Maine Christos, K~Watanabe, T~Taniguchi, Subir Sachdev, Mathias~S Scheurer, Efthimios Kaxiras, Cory~R Dean, et~al.
\newblock Orderly disorder in magic-angle twisted trilayer graphene.
\newblock {\em Science}, 376(6589):193--199, 2022.

\bibitem{wang2006new}
Feng Wang, JV~Alvarez, S-K Mo, JW~Allen, G-H Gweon, J~He, Rongying Jin, David Mandrus, and H~H{\"o}chst.
\newblock New luttinger-liquid physics from photoemission on li 0.9 mo 6 o 17.
\newblock {\em Physical review letters}, 96(19):196403, 2006.

\bibitem{watson2017multiband}
MD~Watson, Y~Feng, CW~Nicholson, Claude Monney, JM~Riley, H~Iwasawa, K~Refson, V~Sacksteder, DT~Adroja, J~Zhao, et~al.
\newblock Multiband one-dimensional electronic structure and spectroscopic signature of tomonaga-luttinger liquid behavior in k 2 cr 3 as 3.
\newblock {\em Physical review letters}, 118(9):097002, 2017.

\bibitem{weston2020atomic}
Astrid Weston, Yichao Zou, Vladimir Enaldiev, Alex Summerfield, Nicholas Clark, Viktor Z{\'o}lyomi, Abigail Graham, Celal Yelgel, Samuel Magorrian, Mingwei Zhou, et~al.
\newblock Atomic reconstruction in twisted bilayers of transition metal dichalcogenides.
\newblock {\em Nature nanotechnology}, 15(7):592--597, 2020.

\bibitem{wickramaratne2015electronic}
Darshana Wickramaratne, Ferdows Zahid, and Roger~K Lake.
\newblock Electronic and thermoelectric properties of van der waals materials with ring-shaped valence bands.
\newblock {\em Journal of Applied Physics}, 118(7), 2015.

\bibitem{wigner1934interaction}
Eugene Wigner.
\newblock On the interaction of electrons in metals.
\newblock {\em Physical Review}, 46(11):1002, 1934.

\bibitem{wu2018hubbard}
Fengcheng Wu, Timothy Lovorn, Emanuel Tutuc, and Allan~H MacDonald.
\newblock Hubbard model physics in transition metal dichalcogenide moir{\'e} bands.
\newblock {\em Physical review letters}, 121(2):026402, 2018.

\bibitem{xia2024unconventional}
Yiyu Xia, Zhongdong Han, Kenji Watanabe, Takashi Taniguchi, Jie Shan, and Kin~Fai Mak.
\newblock {Unconventional superconductivity in twisted bilayer WSe2}.
\newblock {\em arXiv (Cornell University)}, 5 2024.

\bibitem{xu2020correlated}
Yang Xu, Song Liu, Daniel~A. Rhodes, Kenji Watanabe, Takashi Taniguchi, James Hone, Veit Elser, Kin~Fai Mak, and Jie Shan.
\newblock {Correlated insulating states at fractional fillings of moiré superlattices}.
\newblock {\em Nature}, 587(7833):214--218, 11 2020.

\bibitem{yu2023evidence}
Guo Yu, Pengjie Wang, Ayelet~J Uzan-Narovlansky, Yanyu Jia, Michael Onyszczak, Ratnadwip Singha, Xin Gui, Tiancheng Song, Yue Tang, Kenji Watanabe, et~al.
\newblock Evidence for two dimensional anisotropic luttinger liquids at millikelvin temperatures.
\newblock {\em Nature communications}, 14(1):7025, 2023.

\bibitem{zeng2023thermodynamics}
Yihang Zeng, Zhengchao Xia, Kaifei Kang, Jiacheng Zhu, Patrick Knüppel, Chirag Vaswani, Kenji Watanabe, Takashi Taniguchi, Kin~Fai Mak, and Jie Shan.
\newblock {Thermodynamic evidence of fractional Chern insulator in moiré MoTe2}.
\newblock {\em Nature}, 622(7981):69--73, 7 2023.

\bibitem{zhou2021superconductivity}
Haoxin Zhou, Tian Xie, Takashi Taniguchi, Kenji Watanabe, and Andrea~F. Young.
\newblock {Superconductivity in rhombohedral trilayer graphene}.
\newblock {\em Nature}, 598(7881):434--438, 9 2021.

\end{thebibliography}


\begin{thebibliography}{10}

\bibitem{bistritzer2011moire}
Rafi Bistritzer and Allan~H. MacDonald.
\newblock {Moiré bands in twisted double-layer graphene}.
\newblock {\em Proceedings of the National Academy of Sciences}, 108(30):12233--12237, 7 2011.

\bibitem{dajornada2017nonuniform}
Felipe~H. Da~Jornada, Diana~Y. Qiu, and Steven~G. Louie.
\newblock {Nonuniform sampling schemes of the Brillouin zone for many-electron perturbation-theory calculations in reduced dimensionality}.
\newblock {\em Physical review. B./Physical review. B}, 95(3), 1 2017.

\bibitem{deslippe2011berkeley}
Jack Deslippe, Georgy Samsonidze, David~A. Strubbe, Manish Jain, Marvin~L. Cohen, and Steven~G. Louie.
\newblock {BerkeleyGW: A massively parallel computer package for the calculation of the quasiparticle and optical properties of materials and nanostructures}.
\newblock {\em Computer Physics Communications}, 183(6):1269--1289, 12 2011.

\bibitem{garcia2020siestarecent}
Alberto García, Nick Papior, Arsalan Akhtar, Emilio Artacho, Volker Blum, Emanuele Bosoni, Pedro Brandimarte, Mads Brandbyge, J.~I. Cerdá, Fabiano Corsetti, Ramón Cuadrado, Vladimir Dikan, Jaime Ferrer, Julian Gale, Pablo García-Fernández, V.~M. García-Suárez, Sandra García, Georg Huhs, Sergio Illera, Richard Korytár, Peter Koval, Irina Lebedeva, Lin Lin, Pablo López-Tarifa, Sara~G. Mayo, Stephan Mohr, Pablo Ordejón, Andrei Postnikov, Yann Pouillon, Miguel Pruneda, Roberto Robles, Daniel Sánchez-Portal, Jose~M. Soler, Rafi Ullah, Victor Wen-Zhe Yu, and Javier Junquera.
\newblock {Siesta: Recent developments and applications}.
\newblock {\em The Journal of Chemical Physics}, 152(20), 5 2020.

\bibitem{hybertsen1986electron}
Mark~S. Hybertsen and Steven~G. Louie.
\newblock {Electron correlation in semiconductors and insulators: Band gaps and quasiparticle energies}.
\newblock {\em Physical review. B, Condensed matter}, 34(8):5390--5413, 10 1986.

\bibitem{liu2013three}
Gui-Bin Liu, Wen-Yu Shan, Yugui Yao, Wang Yao, and Di~Xiao.
\newblock Three-band tight-binding model for monolayers of group-vib transition metal dichalcogenides.
\newblock {\em Physical Review B—Condensed Matter and Materials Physics}, 88(8):085433, 2013.

\bibitem{liu2018nature}
Hongsheng Liu, Paolo Lazzaroni, and Cristiana Di~Valentin.
\newblock Nature of excitons in bidimensional wse2 by hybrid density functional theory calculations.
\newblock {\em Nanomaterials}, 8(7):481, 2018.

\bibitem{naik2019kolmogorov}
Mit~H Naik, Indrajit Maity, Prabal~K Maiti, and Manish Jain.
\newblock Kolmogorov--crespi potential for multilayer transition-metal dichalcogenides: capturing structural transformations in moir{\'e} superlattices.
\newblock {\em The Journal of Physical Chemistry C}, 123(15):9770--9778, 2019.

\bibitem{naik2021twister}
Saismit Naik, Mit~H. Naik, Indrajit Maity, and Manish Jain.
\newblock {Twister: Construction and structural relaxation of commensurate moiré superlattices}.
\newblock {\em Computer Physics Communications}, 271:108184, 10 2021.

\bibitem{perdew1996generalized}
John~P. Perdew, Kieron Burke, and Matthias Ernzerhof.
\newblock {Generalized gradient approximation made simple}.
\newblock {\em Physical Review Letters}, 77(18):3865--3868, 10 1996.

\bibitem{plimpton1995lammps}
S.~Plimpton.
\newblock Fast parallel algorithms for short-range molecular dynamics.
\newblock {\em J. Comput. Phys.}, 117:1--19, 1995.

\bibitem{soler2002siesta}
José~M Soler, Emilio Artacho, Julian~D Gale, Alberto García, Javier Junquera, Pablo Ordejón, and Daniel Sánchez-Portal.
\newblock {The SIESTA method forab initioorder-Nmaterials simulation}.
\newblock {\em Journal of Physics Condensed Matter}, 14(11):2745--2779, 3 2002.

\bibitem{stillinger1985potential}
F.~H. Stillinger and T.~A. Weber.
\newblock Computer simulation of local order in condensed phases of silicon.
\newblock {\em Phys. Rev. B}, 31:5262--5271, 1985.

\bibitem{thompson2022lammps}
A.~P. Thompson, H.~M. Aktulga, R.~Berger, et~al.
\newblock Lammps - a flexible simulation tool for particle-based materials modeling at the atomic, meso, and continuum scales.
\newblock {\em Comput. Phys. Commun.}, 271:108171, 2022.

\bibitem{van1953occurrence}
L{\'e}on Van~Hove.
\newblock The occurrence of singularities in the elastic frequency distribution of a crystal.
\newblock {\em Physical Review}, 89(6):1189, 1953.

\end{thebibliography}
\end{document}